\documentclass{article}
\usepackage{amssymb}
\usepackage{amsmath}

\allowdisplaybreaks

\begin{document}

\title{On the uniqueness of $D=11$ interactions among a graviton, a massless
gravitino and a three-form. I: Pauli-Fierz and three-form}

\author{E. M. Cioroianu\thanks{ e-mail address:
manache@central.ucv.ro}, E. Diaconu\thanks{ e-mail address:
ediaconu@central.ucv.ro}, S. C. Sararu\thanks{ e-mail
address: scsararu@central.ucv.ro}\\
Faculty of Physics, University of Craiova,\\ 13 Al. I. Cuza Street
Craiova, 200585, Romania}
\date{}
\maketitle

\begin{abstract}
Cross-couplings between a massless spin-two field (described in the free
limit by the Pauli-Fierz action) and an Abelian three-form gauge field in $%
D=11$ are investigated in the framework of the deformation theory based on
local BRST cohomology. These consistent interactions are obtained on the
grounds of smoothness in the coupling constant, locality, Lorentz
covariance, Poincar\'{e} invariance, and the presence of at most two
derivatives in the interacting Lagrangian. Our results confirm the
uniqueness of the eleven-dimensional interactions between a graviton and a
three-form prescribed by General Relativity.

PACS number: 11.10.Ef

\end{abstract}

\section{Introduction}

A key point in the development of the BRST formalism was its cohomological
understanding, which allowed, among others, a useful investigation of many
interesting aspects related to the perturbative renormalization problem~\cite%
{4}--\cite{5}, the anomaly-tracking mechanism~\cite{5}--\cite{6'}, the
simultaneous study of local and rigid invariances of a given theory~\cite{7}
as well as the reformulation of the construction of consistent interactions
in gauge theories~\cite{7a}--\cite{7d} in terms of the deformation theory~%
\cite{8a}--\cite{8d} or, actually, in terms of the deformation of the
solution to the master equation. The impossibility of cross-interactions
among several Einstein (Weyl) gravitons, see Ref.~\cite{multi} (or
respectively Ref.~\cite{marcann}), and of cross-couplings among different
Einstein gravitons in the presence of matter fields~\cite{multi,bizjhep}--%
\cite{bizannalen} has recently been shown by means of cohomological
arguments. In the same context the uniqueness of $D=4$, $N=1$ supergravity
was proved in Ref.~\cite{boulcqg}.

On the other hand, $D=11$, $N=1$ supergravity~\cite{scherk,wit} has regained
a central role with the advent of $M$-theory, whose QFT (local) limit it is.
Of the many special properties of $D=11$, $N=1$ supergravity, one of the
most striking is that it forbids a cosmological term. The proof of this
result has been done in Ref.~\cite{hendessem97} using a combined technique
--- the standard Noether current method and a cohomological approach. It is
known that the field content of $D=11$, $N=1$ supergravity is quite simple;
it comprises a graviton, a massless Majorana spin-$3/2$ field, and a
three-form gauge field. The analysis of all possible interactions in $D=11$
related to this field content necessitates the study of cross-couplings
involving each pair of these sorts of fields and then the construction of
simultaneous interactions among all the three types of fields. One of the
most efficient and meanwhile elegant approaches to the problem of
constructing consistent interactions in gauge field theories\footnote{%
By `consistent' we mean that the interacting theory preserves both the field
content and the number of independent gauge symmetries of the free one.} is
that based on the deformation technique~\cite{8a} combined with local BRST
cohomology~\cite{gen1,gen2}. This approach relies on computing the
deformations of the solution to the master equation for the interacting
theory with the help of the `free' BRST cohomology. Our main aim is to
construct all consistent interactions in $D=11$ that can be added to a free
theory describing a Pauli-Fierz graviton, a massless Rarita-Schwinger
gravitino, and an Abelian three-form gauge field from the deformation of the
`free' solution to the master equation such that the interactions satisfy
some general and quite natural assumptions (smoothness in the coupling
constant, locality, Lorentz covariance, Poincar\'{e} invariance, and
preservation of the differential order of the free field equations at the
level of the coupled theory). One of the final outcomes of this procedure
will be the quest for the uniqueness of $D=11$, $N=1$ SUGRA. In order to
organize the results as logical as possible, to expose in detail the
cohomological aspects involved, and (last but not least) make various
comments on and comparisons with other results from the literature we chose
to split our work into four main parts. The first three are dedicated to the
construction of consistent interactions that involve only two of the three
types of fields under considerations: i) a graviton and a three-form
(present paper); ii) a three-form and massless gravitini~\cite{SUGRAII};
iii) massless gravitini and a graviton~\cite{SUGRAIII}. The fourth and last
part~\cite{SUGRAIV} will put the things together and present what happens
when all these fields are present: what new vertices appear, how consistent
are those obtained from the previous steps, and how does the overall coupled
theory looks like.

In this work we implement the first of the four steps explained in the
above, namely we analyze the cross-couplings between a massless spin-two
field (described in the free limit by the Pauli-Fierz action~\cite{pf,pf1})
and an Abelian three-form gauge field in eleven spacetime dimensions. The
cross-interactions are obtained under the hypotheses of smoothness of the
interactions in the coupling constant, locality, Poincar\'{e} invariance,
Lorentz covariance, and the presence of at most two derivatives in the
Lagrangian of the interacting theory (the same number of derivatives like in
the free Lagrangian). Our results are obtained in the context of the
deformation of the solution to the master equation.

We compute the interaction terms to order two in the coupling constant. In
this way we obtain that the first two orders of the interacting Lagrangian
resulting from our setting originate in the development of the full
interacting Lagrangian (in eleven spacetime dimensions)
\begin{equation*}
\tilde{\mathcal{L}}=\frac{2}{\lambda ^{2}}\sqrt{g}\left( R-2\lambda
^{2}\Lambda \right) +\mathcal{L}^{\mathrm{h-A}},
\end{equation*}%
where the cross-coupling part reads as%
\begin{equation*}
\mathcal{L}^{\mathrm{h-A}}=-\frac{1}{2\cdot 4!}\sqrt{g}\bar{F}_{\mu
\nu \rho \lambda }\bar{F}^{\mu \nu \rho \lambda }+\lambda q\epsilon
^{\mu _{1}\ldots \mu _{11}}\bar{A}_{\mu _{1}\mu _{2}\mu
_{3}}\bar{F}_{\mu _{4}\ldots \mu _{7}}\bar{F}_{\mu _{8}\ldots \mu
_{11}},
\end{equation*}%
with $g=\det g_{\mu \nu }$, $\Lambda $ the cosmological constant, $\lambda $
the coupling constant, and $q$ an arbitrary, real constant. Consequently, we
show the uniqueness of interactions described by $\tilde{\mathcal{L}}$. The
above interacting Lagrangian for $\Lambda =0$ is a part of $D=11$, $N=1$
SUGRA Lagrangian. We note that the graviton sector is allowed at this stage
to include a cosmological term, unlike $D=11$, $N=1$ SUGRA. This is not a
surprise since it is the simultaneous presence of all fields (supplemented
with massless gravitini) that ensures the annihilation of the cosmological
constant, as it will be made clear in Ref.~\cite{SUGRAIV}.

This paper is organized in six sections. In section \ref{free} we construct
the BRST symmetry of the free model, consisting in a Pauli-Fierz and an
Abelian three-form gauge field. Section \ref{brief} briefly addresses the
deformation procedure based on BRST symmetry. In section \ref{inter} we
compute the first two orders of the interactions between the massless
spin-two field and an Abelian three-form gauge field. Section \ref{defth} is
devoted to analyzing the deformed theory obtained in the previous section.
In this context we obtain a possible candidate that describes the
interacting theory to all orders in the coupling constant. Section \ref%
{unique} is dedicated to the investigation of the uniqueness of interactions
described by the candidate emphasized in the previous section. The last
section exposes the main conclusions on this paper.

\section{Free model: Lagrangian formulation and BRST symmetry\label{free}}

Our starting point is represented by a free Lagrangian action, written as
the sum between the linearized Hilbert-Einstein action (also known as the
Pauli-Fierz action) and the action for an Abelian three-form gauge field in
eleven spacetime dimensions
\begin{eqnarray}
S_{0}^{\mathrm{L}}\left[ h_{\mu \nu },A_{\mu \nu \rho }\right] &=&\int
d^{11}x\left( -\frac{1}{2}\left( \partial _{\mu }h_{\nu \rho }\right) \left(
\partial ^{\mu }h^{\nu \rho }\right) +\left( \partial _{\mu }h^{\mu \rho
}\right) \left( \partial ^{\nu }h_{\nu \rho }\right) \right.  \notag \\
&&\left. -\left( \partial _{\mu }h\right) \left( \partial _{\nu }h^{\nu \mu
}\right) +\frac{1}{2}\left( \partial _{\mu }h\right) \left( \partial ^{\mu
}h\right) -\frac{1}{2\cdot 4!}F_{\mu \nu \rho \lambda }F^{\mu \nu \rho
\lambda }\right)  \notag \\
&\equiv &\int d^{11}x\left( \mathcal{L}^{\mathrm{h}}+\mathcal{L}_{0}^{%
\mathrm{A}}\right) .  \label{fract}
\end{eqnarray}%
Throughout the paper we work with the flat metric of `mostly minus'
signature, $\sigma _{\mu \nu }=\left( +-\cdots -\right) $. In the above $h$
denotes the trace of the Pauli-Fierz field, $h=\sigma _{\mu \nu }h^{\mu \nu
} $, and $F_{\mu \nu \rho \lambda }$ denotes the field-strength of the
three-form gauge field ($F_{\mu \nu \rho \lambda }\equiv \partial _{\lbrack
\mu }A_{\nu \rho \lambda ]}$). The notation $\left[ \mu \ldots \nu \right] $
(respectively $\left( \mu \ldots \nu \right) $) signifies antisymmetry
(respectively symmetry) with respect to all indices between brackets without
normalization factors (i.e., the independent terms appear only once and are
not multiplied by overall numerical factors). The theory described by action
(\ref{fract}) possesses an Abelian generating set of gauge transformations
\begin{equation}
\delta _{\epsilon ,\varepsilon }h_{\mu \nu }=\partial _{(\mu }\epsilon _{\nu
)},\qquad \delta _{\epsilon ,\varepsilon }A_{\mu \nu \rho }=\partial
_{\lbrack \mu }\varepsilon _{\nu \rho ]},  \label{f2}
\end{equation}%
where the gauge parameters $\epsilon ^{\Gamma _{1}}\equiv \left\{ \epsilon
_{\mu },\varepsilon _{\mu \nu }\right\} $ are bosonic functions, with the
last set completely antisymmetric. We observe that if in (\ref{f2}) we make
the transformations
\begin{equation}
\varepsilon _{\mu \nu }\rightarrow \varepsilon _{\mu \nu }^{\left( \theta
\right) }=\partial _{\lbrack \mu }\theta _{\nu ]},  \label{f3}
\end{equation}%
then the gauge variation of the three-form identically vanishes
\begin{equation}
\delta _{\varepsilon ^{\left( \theta \right) }}A_{\mu \nu \rho }\equiv 0.
\label{f3a}
\end{equation}%
Moreover, if in (\ref{f3}) we perform the changes
\begin{equation}
\theta _{\mu }\rightarrow \theta _{\mu }^{\left( \phi \right) }=\partial
_{\mu }\phi ,  \label{f4}
\end{equation}%
with $\phi $\ an arbitrary scalar field, then the transformed gauge
parameters from (\ref{f3}) identically vanish
\begin{equation}
\varepsilon _{\mu \nu }^{\left( \theta ^{\left( \phi \right) }\right)
}\equiv 0.  \label{f4a}
\end{equation}%
Meanwhile, there is no nonvanishing local transformation of $\phi $\ that
annihilates $\theta _{\mu }^{\left( \phi \right) }$\ of the form (\ref{f4}),
and hence no further local reducibility identity. All these allow us to
conclude that the generating set of gauge transformations given in (\ref{f2}%
) is off-shell, second-stage reducible. It is obvious that the accompanying
gauge algebra is Abelian.

In order to construct the BRST symmetry for (\ref{fract}) we introduce the
field, ghost, and antifield spectra
\begin{eqnarray}
&\Phi ^{\Gamma _{0}}=\left( h_{\mu \nu },A_{\mu \nu \rho }\right) ,\qquad
&\Phi _{\Gamma _{0}}^{\ast }=\left( h^{\ast \mu \nu },A^{\ast \mu \nu \rho
}\right)  \label{f6a} \\
&\eta ^{\Gamma _{1}}=\left( \eta _{\mu },C_{\mu \nu }\right) ,\qquad &\eta
_{\Gamma _{1}}^{\ast }=\left( \eta ^{\ast \mu },C^{\ast \mu \nu }\right) ,
\label{f6b} \\
&\eta ^{\Gamma _{2}}=\left( C_{\mu }\right) ,\qquad &\eta _{\Gamma
_{2}}^{\ast }=\left( C^{\ast \mu }\right) ,  \label{f6c} \\
&\eta ^{\Gamma _{3}}=\left( C\right) ,\qquad &\eta _{\Gamma _{3}}^{\ast
}=\left( C^{\ast }\right) .  \label{f6d}
\end{eqnarray}%
The fermionic ghosts $\eta ^{\Gamma _{1}}$ respectively correspond to the
bosonic gauge parameters $\epsilon ^{\Gamma _{1}}$ from (\ref{f2}), the
bosonic ghosts for ghosts $\eta ^{\Gamma _{2}}$ are associated with the
first-stage reducibility parameters $\theta _{\mu }$ in (\ref{f3}), while
the fermionic ghost for ghost for ghost $\eta ^{\Gamma _{3}}$ is present due
to the second-stage reducibility parameter $\phi $ from (\ref{f4}). The star
variables represent the antifields of the corresponding fields/ghosts. Their
Grassmann parities are obtained via the standard rule of the BRST method $%
\varepsilon \left( \chi _{\Gamma }^{\ast }\right) =\left( \varepsilon \left(
\chi ^{\Gamma }\right) +1\right) \,\mathrm{mod\,}2$, where we employed the
notations
\begin{equation}
\chi ^{\Gamma }=\left( \Phi ^{\Gamma _{0}},\eta ^{\Gamma _{1}},\eta ^{\Gamma
_{2}},\eta ^{\Gamma _{3}}\right) ,\qquad \chi _{\Gamma }^{\ast }=\left( \Phi
_{\Gamma _{0}}^{\ast },\eta _{\Gamma _{1}}^{\ast },\eta _{\Gamma _{2}}^{\ast
},\eta _{\Gamma _{3}}^{\ast }\right) .  \label{notat}
\end{equation}

Since both the gauge generators and the reducibility functions for this
model are field-independent, it follows that the BRST differential $s$
reduces to
\begin{equation}
s=\delta +\gamma ,  \label{desc}
\end{equation}%
where $\delta $ is the Koszul-Tate differential and $\gamma $ denotes the
exterior longitudinal derivative. The Koszul-Tate differential is graded in
terms of the antighost number ($\mathrm{agh}$, $\mathrm{agh}\left( \delta
\right) =-1$, $\mathrm{agh}\left( \gamma \right) =0$) and enforces a
resolution of the algebra of smooth functions defined on the stationary
surface of field equations for action (\ref{fract}), $C^{\infty }\left(
\Sigma \right) $, $\Sigma :\delta S_{0}^{\mathrm{L}}/\delta \Phi ^{\alpha
_{0}}=0$. The exterior longitudinal derivative is graded in terms of the
pure ghost number ($\mathrm{pgh}$, $\mathrm{pgh}\left( \gamma \right) =1$, $%
\mathrm{pgh}\left( \delta \right) =0$) and is correlated with the original
gauge symmetry via its cohomology in pure ghost number zero computed in $%
C^{\infty }\left( \Sigma \right) $, which is isomorphic to the algebra of
physical observables for this free theory. These two degrees of the
generators (\ref{f6a})--(\ref{f6d}) from the BRST complex are valued as
\begin{eqnarray}
&\mathrm{pgh}\left( \Phi ^{\Gamma _{0}}\right) =0,\qquad &\mathrm{pgh}\left(
\eta ^{\Gamma _{k}}\right) =k,  \label{f7c} \\
&\mathrm{pgh}\left( \Phi _{\Gamma _{0}}^{\ast }\right) =0,\qquad &\mathrm{pgh%
}\left( \eta _{\Gamma _{k}}^{\ast }\right) =0,  \label{f7a} \\
&\mathrm{agh}\left( \Phi ^{\Gamma _{0}}\right) =0,\qquad &\mathrm{agh}\left(
\eta ^{\Gamma _{k}}\right) =0,  \label{f7b} \\
&\mathrm{agh}\left( \Phi _{\Gamma _{0}}^{\ast }\right) =1,\qquad &\mathrm{agh%
}\left( \eta _{\Gamma _{k}}^{\ast }\right) =k+1,  \label{f7d}
\end{eqnarray}%
for $k=\overline{1,3}$. The actions of the differentials $\delta $ and $%
\gamma $ on the generators from the BRST complex are given by
\begin{eqnarray}
&&\delta h^{\ast \mu \nu }=2H^{\mu \nu },\qquad \delta A^{\ast \mu \nu \rho
}=\frac{1}{3!}\partial _{\lambda }F^{\mu \nu \rho \lambda },  \label{f8a} \\
&&\delta \eta ^{\ast \mu }=-2\partial _{\nu }h^{\ast \mu \nu },\qquad \delta
C^{\ast \mu \nu }=-3\partial _{\rho }A^{\ast \mu \nu \rho },  \label{f8b} \\
&&\delta C^{\ast \mu }=-2\partial _{\nu }C^{\ast \mu \nu },\qquad \delta
C^{\ast }=-\partial _{\mu }C^{\ast \mu },\qquad \delta \chi ^{\Gamma }=0,
\label{f8d} \\
&&\gamma \chi _{\Gamma }^{\ast }=0,\qquad \gamma h_{\mu \nu }=\partial
_{(\mu }\eta _{\nu )},\qquad \gamma A_{\mu \nu \rho }=\partial _{\lbrack \mu
}C_{\nu \rho ]},  \label{f8f} \\
&&\gamma \eta _{\mu }=0,\qquad \gamma C_{\mu \nu }=\partial _{\lbrack \mu
}C_{\nu ]},\qquad \gamma C_{\mu }=\partial _{\mu }C,\qquad \gamma C=0.
\label{f8h}
\end{eqnarray}%
In the above $H^{\mu \nu }=K^{\mu \nu }-\frac{1}{2}\sigma ^{\mu \nu }K$ is
the linearized Einstein tensor, with $K^{\mu \nu }$ and $K$ the linearized
Ricci tensor and respectively the linearized scalar curvature, both obtained
from the linearized Riemann tensor $K_{\mu \nu \alpha \beta }=\frac{1}{2}%
\partial _{\lbrack \mu }h_{\nu ][\alpha ,\beta ]}$\ via its trace and
respectively double trace: $K_{\mu \alpha }=\sigma ^{\nu \beta }K_{\mu \nu
\alpha \beta }$ and respectively $K=\sigma ^{\mu \alpha }\sigma ^{\nu \beta
}K_{\mu \nu \alpha \beta }.$

The BRST differential is known to have a canonical action in a
structure named antibracket and denoted by the symbol $\left(
,\right) $ ($s\cdot =\left( \cdot ,S\right) $), which is obtained by
considering the fields/ghosts respectively conjugated to the
corresponding antifields. The generator of the BRST symmetry is a
bosonic functional of ghost number zero,
which is solution to the classical master equation $\left( S,S%
\right) =0$. The full solution to the master equation for the free model
under study reads as
\begin{equation}
S^{\mathrm{h,A}}=S_{0}^{\mathrm{L}}+\int d^{11}x\left( h^{\ast \mu \nu
}\partial _{(\mu }\eta _{\nu )}+A^{\ast \mu \nu \rho }\partial _{\lbrack \mu
}C_{\nu \rho ]}+C^{\ast \mu \nu }\partial _{\lbrack \mu }C_{\nu ]}+C^{\ast
\mu }\partial _{\mu }C\right) .  \label{f12}
\end{equation}%
The solution to the master equation encodes all the information on the gauge
structure of a given theory.

\section{Deformation of the solution to the master equation: a brief review
\label{brief}}

We begin with a ``free'' gauge theory, described by a Lagrangian action $%
S_{0}^{\mathrm{L}}\left[ \Phi ^{\Gamma _{0}}\right] $, invariant under some
gauge transformations $\delta _{\epsilon }\Phi ^{\Gamma _{0}}=Z_{\;\;\Gamma
_{1}}^{\Gamma _{0}}\epsilon ^{\Gamma _{1}}$, i.e. $\frac{\delta S_{0}^{%
\mathrm{L}}}{\delta \Phi ^{\Gamma _{0}}}Z_{\;\;\Gamma _{1}}^{\Gamma _{0}}=0$%
, and consider the problem of constructing consistent interactions among the
fields $\Phi ^{\Gamma _{0}}$ such that the couplings preserve the field
spectrum and the original number of gauge symmetries. This matter is
addressed by means of reformulating the problem of constructing consistent
interactions as a deformation problem of the solution to the master equation
corresponding to the ``free'' theory~\cite{8a}. Such a reformulation is
possible due to the fact that the solution to the master equation contains
all the information on the gauge structure of the theory. If an interacting
gauge theory can be consistently constructed, then the solution $S$ to the
master equation associated with the ``free'' theory, $\left( S,S\right) =0$,
can be deformed into a solution $\bar{S}$
\begin{equation}
S\rightarrow \bar{S}=S+\lambda S_{1}+\lambda ^{2}S_{2}+\cdots =S+\lambda
\int d^{D}x\,a+\lambda ^{2}\int d^{D}x\,b+\cdots  \label{2.2}
\end{equation}
of the master equation for the deformed theory
\begin{equation}
\left( \bar{S},\bar{S}\right) =0,  \label{2.3}
\end{equation}
such that both the ghost and antifield spectra of the initial theory are
preserved. Equation (\ref{2.3}) splits, according to the various orders in
the coupling constant (deformation parameter) $\lambda $, into a tower of
equations:
\begin{eqnarray}
\left( S,S\right) &=&0  \label{2.4} \\
2\left( S_{1},S\right) &=&0  \label{2.5} \\
2\left( S_{2},S\right) +\left( S_{1},S_{1}\right) &=&0  \label{2.6} \\
\left( S_{3},S\right) +\left( S_{1},S_{2}\right) &=&0  \label{2.7} \\
&&\vdots  \notag
\end{eqnarray}

Equation (\ref{2.4}) is fulfilled by hypothesis. The next equation requires
that the first-order deformation of the solution to the master equation, $%
S_{1}$, is a co-cycle of the \textquotedblleft free\textquotedblright\ BRST
differential $s$, $sS_{1}=0$. However, only cohomologically nontrivial
solutions to (\ref{2.5}) should be taken into account, since the BRST-exact
ones can be eliminated by some (in general nonlinear) field redefinitions.
This means that $S_{1}$ pertains to the ghost number zero cohomological
space of $s$, $H^{0}\left( s\right) $, which is generically nonempty because
it is isomorphic to the space of physical observables of the
\textquotedblleft free\textquotedblright\ theory. It has been shown (by of
the triviality of the antibracket map in the cohomology of the BRST
differential) that there are no obstructions in finding solutions to the
remaining equations, namely (\ref{2.6})--(\ref{2.7}), etc. However, the
resulting interactions may be nonlocal and there might even appear
obstructions if one insists on their locality. The analysis of these
obstructions can be done with the help of cohomological techniques.

\section{Consistent interactions between the Pauli-Fierz field and an
Abelian three-form gauge field\label{inter}}

\subsection{Standard material: basic cohomologies\label{stand}}

The aim of this section is to investigate the cross-couplings that can be
introduced between a Pauli-Fierz field and an Abelian three-form gauge
field. This matter is addressed in the context of the antifield-BRST
deformation procedure described in the above and relies on computing the
solutions to equations (\ref{2.5})--(\ref{2.7}), etc., with the help of the
BRST cohomology of the free theory. The interactions are obtained under the
following (reasonable) assumptions: smoothness in the deformation parameter,
locality, Lorentz covariance, Poincar\'{e} invariance, and the presence of
at most two derivatives in the interacting Lagrangian. `Smoothness in the
deformation parameter' refers to the fact that the deformed solution to the
master equation, (\ref{2.2}), is smooth in the coupling constant $\lambda $
and reduces to the original solution, (\ref{f12}), in the free limit $%
\lambda =0$. The requirement on the interacting theory to be Poincar\'{e}
invariant means that one does not allow an explicit dependence on the
spacetime coordinates into the deformed solution to the master equation. The
requirement concerning the maximum number of derivatives allowed to enter
the interacting Lagrangian is frequently imposed in the literature at the
level of interacting theories; for instance, see the case of
cross-interactions for a collection of Pauli-Fierz fields, Ref.~\cite{multi}%
, the couplings between the Pauli-Fierz and the massless Rarita-Schwinger
fields, Ref.~\cite{boulcqg}, or the direct cross-interactions for a
collection of Weyl gravitons, Ref.~\cite{marcann}. Equation (\ref{2.5}),
which we have seen that controls the first-order deformation, takes the
local form
\begin{equation}
sa=\partial _{\mu }m^{\mu },\qquad \mathrm{gh}\left( a\right) =0,\qquad
\varepsilon \left( a\right) =0,  \label{3.1}
\end{equation}%
for some local $m^{\mu }$, and it shows that the nonintegrated density of
the first-order deformation pertains to the local cohomology of the free
BRST\ differential in ghost number zero, $a\in H^{0}\left( s|d\right) $,
where $d$ denotes the exterior spacetime differential. The solution to (\ref%
{3.1}) is unique up to $s$-exact pieces plus divergences
\begin{equation}
a\rightarrow a+sb+\partial _{\mu }n^{\mu },  \label{3.1a}
\end{equation}%
with $\mathrm{gh}\left( b\right) =-1$, $\varepsilon \left( b\right) =1$, $%
\mathrm{gh}\left( n^{\mu }\right) =0$, and $\varepsilon \left( n^{\mu
}\right) =0$. At the same time, if the general solution of (\ref{3.1}) is
found to be completely trivial, $a=sb+\partial _{\mu }n^{\mu }$, then it can
be made to vanish $a=0$.

In order to analyze equation (\ref{3.1}), we develop $a$ according to the
antighost number
\begin{equation}
a=\sum\limits_{i=0}^{I}a_{i},\qquad \mathrm{agh}\left( a_{i}\right)
=i,\qquad \mathrm{gh}\left( a_{i}\right) =0,\qquad \varepsilon \left(
a_{i}\right) =0,  \label{3.2}
\end{equation}%
and assume, without loss of generality, that decomposition (\ref{3.2}) stops
at some finite value of $I$. This can be shown for instance like in Appendix
A of Ref.~\cite{multi}. Replacing decomposition (\ref{3.2}) into (\ref{3.1})
and projecting it on the various values of the antighost number by means of (%
\ref{desc}), we obtain the tower of equations
\begin{eqnarray}
\gamma a_{I} &=&\partial _{\mu }\overset{\left( I\right) }{m}^{\mu },
\label{3.3} \\
\delta a_{I}+\gamma a_{I-1} &=&\partial _{\mu }\overset{\left( I-1\right) }{m%
}^{\mu },  \label{3.4} \\
\delta a_{i}+\gamma a_{i-1} &=&\partial _{\mu }\overset{\left( i-1\right) }{m%
}^{\mu },\qquad 1\leq i\leq I-1,  \label{3.5}
\end{eqnarray}%
where $\left( \overset{\left( i\right) }{m}^{\mu }\right) _{i=\overline{0,I}%
} $ are some local currents, with $\mathrm{agh}\left( \overset{\left(
i\right) }{m}^{\mu }\right) =i$. Moreover, according to the general result
from Ref.~\cite{multi} in the absence of collection indices, equation (\ref%
{3.3}) can be replaced in strictly positive antighost numbers by
\begin{equation}
\gamma a_{I}=0,\qquad I>0.  \label{3.6}
\end{equation}%
Due to the second-order nilpotency of $\gamma $ ($\gamma ^{2}=0$), the
solution to (\ref{3.6}) is unique up to $\gamma $-exact contributions
\begin{equation}
a_{I}\rightarrow a_{I}+\gamma b_{I},\qquad \mathrm{agh}\left( b_{I}\right)
=I,\qquad \mathrm{pgh}\left( b_{I}\right) =I-1,\qquad \varepsilon \left(
b_{I}\right) =1.  \label{r68}
\end{equation}%
Meanwhile, if it turns out that $a_{I}$ reduces to $\gamma $-exact terms
only, $a_{I}=\gamma b_{I}$, then it can be made to vanish, $a_{I}=0$. In
other words, the nontriviality of the first-order deformation $a$ is
translated at its highest antighost number component into the requirement
that $a_{I}\in H^{I}\left( \gamma \right) $, where $H^{I}\left( \gamma
\right) $ denotes the cohomology of the exterior longitudinal derivative $%
\gamma $ in pure ghost number equal to $I$. So, in order to solve equation (%
\ref{3.1}) (equivalent with (\ref{3.6}) and (\ref{3.4})--(\ref{3.5})), we
need to compute the cohomology of $\gamma $, $H\left( \gamma \right) $, and,
as it will be made clear below, also the local cohomology of $\delta $, $%
H\left( \delta |d\right) $.

Using the results on the cohomology of $\gamma $ in the Pauli-Fierz sector~%
\cite{multi} as well as definitions (\ref{f8f}) and (\ref{f8h}), we can
state that $H\left( \gamma \right) $ is generated on the one hand by $\chi
_{\Gamma }^{\ast }$, $F_{\mu \nu \rho \lambda }$, and $K_{\mu \nu \alpha
\beta }$, together with their spacetime derivatives and, on the other hand,
by the undifferentiated ghost for ghost for ghost $C$ as well by the ghosts $%
\eta _{\mu }$ and their first-order derivatives $\partial _{\lbrack \mu
}\eta _{\nu ]}$. So, the most general (and nontrivial) solution to (\ref{3.6}%
) can be written, up to $\gamma $-exact contributions, as
\begin{equation}
a_{I}^{\mathrm{h,A}}=\alpha _{I}\left( \left[ F_{\mu \nu \rho \lambda }%
\right] ,\left[ K_{\mu \nu \alpha \beta }\right] ,\left[ \chi _{\Delta
}^{\ast }\right] \right) \omega ^{I}\left( C,\eta _{\mu },\partial _{\lbrack
\mu }\eta _{\nu ]}\right) ,  \label{3.10}
\end{equation}%
where the notation $f\left( \left[ q\right] \right) $ means that $f$ depends
on $q$ and its derivatives up to a finite order, while $\omega ^{I}$ denotes
the elements of a basis in the space of polynomials with pure ghost number $%
I $ in the corresponding ghost for ghost for ghost, Pauli-Fierz ghosts and
their antisymmetrized first-order derivatives. The objects $\alpha _{I}$
(obviously nontrivial in $H^{0}\left( \gamma \right) $) were taken to have a
finite antighost number and a bounded number of derivatives, and therefore
they are polynomials in the antifields $\chi _{\Gamma }^{\ast }$, in the
linearized Riemann tensor $K_{\mu \nu \alpha \beta }$ and in the
field-strength of the three-form $F_{\mu \nu \rho \lambda }$ as well as in
their subsequent derivatives. They are required to fulfill the property $%
\mathrm{agh}\left( \alpha _{I}\right) =I$ in order to ensure that the ghost
number of $a_{I}$ is equal to zero. Due to their $\gamma $-closeness, $%
\gamma \alpha _{I}=0$, and to their polynomial character, $\alpha _{I}$ will
be called invariant polynomials. In antighost number equal to zero the
invariant polynomials are polynomials in the linearized Riemann tensor, in
the field-strength of the Abelian three-form, and in their derivatives.

Inserting (\ref{3.10}) in (\ref{3.4}), we obtain that a necessary (but not
sufficient) condition for the existence of (nontrivial) solutions $a_{I-1}$
is that the invariant polynomials $\alpha _{I}$ are (nontrivial) objects
from the local cohomology of the Koszul-Tate differential $H\left( \delta
|d\right) $ in antighost number $I>0$ and in pure ghost number zero,
\begin{equation}
\delta \alpha _{I}=\partial _{\mu }\overset{\left( I-1\right) }{j}^{\mu
},\qquad \mathrm{agh}\left( \overset{\left( I-1\right) }{j}^{\mu }\right)
=I-1,\qquad \mathrm{pgh}\left( \overset{\left( I-1\right) }{j}^{\mu }\right)
=0.  \label{3.10a}
\end{equation}%
We recall that the local cohomology $H\left( \delta |d\right) $ is
completely trivial in both strictly positive antighost \textit{and} pure
ghost numbers (for instance, see Theorem 5.4 from Ref.~\cite{gen1} and also
Ref.~\cite{gen2}). Using the fact that the Cauchy order of the free theory
under study is equal to four, the general results from Refs.~\cite{gen1,gen2}%
, according to which the local cohomology of the Koszul-Tate differential in
pure ghost number zero is trivial in antighost numbers strictly greater than
its Cauchy order, ensure that
\begin{equation}
H_{J}\left( \delta |d\right) =0,\qquad J>4,  \label{3.11}
\end{equation}%
where $H_{J}\left( \delta |d\right) $ denotes the local cohomology of the
Koszul-Tate differential in antighost number $J$ and in pure ghost number
zero. It can be shown that any invariant polynomial that is trivial in $%
H_{J}\left( \delta |d\right) $ with $J\geq 4$ can be taken to be trivial
also in $H_{J}^{\mathrm{inv}}\left( \delta |d\right) $. ($H_{J}^{\mathrm{inv}%
}\left( \delta |d\right) $ denotes the invariant characteristic cohomology
in antighost number $J$ --- the local cohomology of the Koszul-Tate
differential in the space of invariant polynomials.) Thus:
\begin{equation}
\left( \alpha _{J}=\delta b_{J+1}+\partial _{\mu }\overset{(J)}{c}^{\mu },\
\mathrm{agh}\left( \alpha _{J}\right) =J\geq 4\right) \Rightarrow \alpha
_{J}=\delta \beta _{J+1}+\partial _{\mu }\overset{(J)}{\gamma }^{\mu },
\label{3.12ax}
\end{equation}%
with both $\beta _{J+1}$ and $\overset{(J)}{\gamma }^{\mu }$ invariant
polynomials. Results (\ref{3.11}) and (\ref{3.12ax}) yield the conclusion
that
\begin{equation}
H_{J}^{\mathrm{inv}}\left( \delta |d\right) =0,\qquad J>4.  \label{3.12x}
\end{equation}%
By proceeding in the same manner like in Refs.~\cite{multi} and~\cite{knaep1}%
, it can be proved that the spaces $\left( H_{J}\left( \delta |d\right)
\right) _{J\geq 2}$ and $\left( H_{J}^{\mathrm{inv}}\left( \delta |d\right)
\right) _{J\geq 2}$ are spanned by
\begin{eqnarray}
&&H_{4}\left( \delta |d\right) ,H_{4}^{\mathrm{inv}}\left( \delta |d\right)
:\quad \left( C^{\ast }\right) ,  \label{omd4} \\
&&H_{3}\left( \delta |d\right) ,H_{3}^{\mathrm{inv}}\left( \delta |d\right)
:\quad \left( C^{\ast \mu }\right) ,  \label{omd3} \\
&&H_{2}\left( \delta |d\right) ,H_{2}^{\mathrm{inv}}\left( \delta |d\right)
:\quad \left( C^{\ast \mu \nu },\eta ^{\ast \mu }\right) .  \label{omd2}
\end{eqnarray}%
In contrast to the groups $\left( H_{J}\left( \delta |d\right) \right)
_{J\geq 2}$ and $\left( H_{J}^{\mathrm{inv}}\left( \delta |d\right) \right)
_{J\geq 2}$, which are finite-dimensional, the cohomology $H_{1}\left(
\delta |d\right) $ in pure ghost number zero, known to be related to global
symmetries and ordinary conservation laws, is infinite-dimensional since the
theory is free. Fortunately, it will not be needed in the sequel.

The previous results on $H\left( \delta |d\right) $ and $H^{\mathrm{inv}%
}\left( \delta |d\right) $ in strictly positive antighost numbers are
important because they control the obstructions to removing the antifields
from the first-order deformation. Based on formulas (\ref{3.11})--(\ref%
{3.12x}), one can successively eliminate all the pieces of antighost number
strictly greater than four from the nonintegrated density of the first-order
deformation by adding only trivial terms. Consequently, one can take
(without loss of nontrivial objects) $I\leq 4$ into the decomposition (\ref%
{3.2}). (The proof of this statement can be realized like in Appendix C from
Ref.~\cite{t31}.) In addition, the last representative reads as in (\ref%
{3.10}), where the invariant polynomial is necessarily a nontrivial object
from $\left( H_{J}^{\mathrm{inv}}\left( \delta |d\right) \right) _{2\leq
J\leq 4}$ or from $H_{1}\left( \delta |d\right) $ for $J=1$.

\subsection{First-order deformation\label{firstord}}

Assuming $I=4$, the nonintegrated density of the first-order deformation, (%
\ref{3.2}), becomes
\begin{equation}
a^{\mathrm{h,A}}=a_{0}^{\mathrm{h,A}}+a_{1}^{\mathrm{h,A}}+a_{2}^{\mathrm{h,A%
}}+a_{3}^{\mathrm{h,A}}+a_{4}^{\mathrm{h,A}}.  \label{3.12}
\end{equation}%
We can further decompose $a$ in a natural manner as
\begin{equation}
a^{\mathrm{h,A}}=a^{\mathrm{h}}+a^{\mathrm{h-A}}+a^{\mathrm{A}},
\label{3.12a}
\end{equation}%
where $a^{\mathrm{h}}$ contains only fields/ghosts/antifields from the
Pauli-Fierz sector, $a^{\mathrm{h-A}}$ describes the cross-interactions
between the two theories (so it effectively mixes both sectors), and $a^{%
\mathrm{A}}$ involves only the three-form gauge field sector. The component $%
a^{\mathrm{h}}$ is completely known~\cite{multi} and individually satisfies
an equation of the type (\ref{3.1}). It admits a decomposition similar to (%
\ref{3.12})
\begin{equation}
a^{\mathrm{h}}=a_{0}^{\mathrm{h}}+a_{1}^{\mathrm{h}}+a_{2}^{\mathrm{h}},
\label{3.12w}
\end{equation}%
where
\begin{eqnarray}
a_{2}^{\mathrm{h}} &=&\frac{1}{2}\eta ^{\ast \mu }\eta ^{\nu }\partial _{%
\left[ \mu \right. }\eta _{\left. \nu \right] },  \label{fo1a} \\
a_{1}^{\mathrm{h}} &=&h^{\ast \mu \rho }\left( \left( \partial _{\rho }\eta
^{\nu }\right) h_{\mu \nu }-\eta ^{\nu }\partial _{\lbrack \mu }h_{\nu ]\rho
}\right) ,  \label{fo1b}
\end{eqnarray}%
and $a_{0}^{\mathrm{h}}$ is the cubic vertex of the Einstein-Hilbert
Lagrangian plus a cosmological term\footnote{%
The terms $a_{2}^{\mathrm{h}}$ and $a_{1}^{\mathrm{h}}$ given in (\ref{fo1a}%
) and (\ref{fo1b}) differ from the corresponding ones in Ref.~\cite{multi}
by a $\gamma $-exact and respectively a $\delta $-exact contribution.
However, the difference between our $a_{2}^{\mathrm{h}}+$ $a_{1}^{\mathrm{h}%
} $ and the corresponding sum from Ref.~\cite{multi} is a $s$-exact modulo $d
$ quantity. Consequently, the associated component of antighost number $0$, $%
a_{0}^{\mathrm{h}}$, is nevertheless the same in both formulations. Thus,
the object $a^{\mathrm{h}}$ and the first-order deformation from Ref.~\cite%
{multi} belong to the same cohomological class from $H^{0}\left( s|d\right) $%
.}%
\begin{equation*}
a_{0}^{\mathrm{h}}=a_{0}^{\mathrm{h-cubic}}-2\Lambda h,
\end{equation*}
with $\Lambda $ the cosmological constant. Due to the fact that $a^{\mathrm{%
h-A}}$ and $a^{\mathrm{A}}$ contain different sorts of fields, it follows
that they are subject to two separate equations
\begin{eqnarray}
sa^{\mathrm{A}} &=&\partial ^{\mu }m_{\mu }^{\mathrm{A}},  \label{fo2a} \\
sa^{\mathrm{h-A}} &=&\partial ^{\mu }m_{\mu }^{\mathrm{h-A}},  \label{fo2b}
\end{eqnarray}%
for some local $m_{\mu }$'s. In the sequel we analyze the general solutions
to these equations. The nontrivial solution $a^{\mathrm{A}}$ to (\ref{fo2a})
is
\begin{equation}
a^{\mathrm{A}}=q\varepsilon ^{\mu _{1}\ldots \mu _{11}}A_{\mu _{1}\mu
_{2}\mu _{3}}F_{\mu _{4}\ldots \mu _{7}}F_{\mu _{8}\ldots \mu _{11}},
\label{fo5}
\end{equation}%
where $q$ is an arbitrary, real constant (for more details, see Ref.~\cite%
{knaep}). In the sequel we analyze the general solution to equation (\ref%
{fo2b}).

In agreement with (\ref{3.12}), we can assume that the solution to (\ref%
{fo2b}) stops at antighost number four ($I=4$)
\begin{equation}
a^{\mathrm{h-A}}=a_{0}^{\mathrm{h-A}}+a_{1}^{\mathrm{h-A}}+a_{2}^{\mathrm{h-A%
}}+a_{3}^{\mathrm{h-A}}+a_{4}^{\mathrm{h-A}},  \label{fo7}
\end{equation}%
where the components on the right-hand side of (\ref{fo7}) are subject to
equations (\ref{3.6}) and (\ref{3.4})--(\ref{3.5}) for $I=4$. Because $%
\alpha _{4}$ is of the type $fC^{\ast }$ ($f$ is an arbitrary constant) and $%
\omega ^{4}(C,\eta _{\mu },\partial _{\lbrack \mu }\eta _{\nu ]})$ is
spanned by
\begin{eqnarray}
&&\left\{ C\eta _{\mu },C\partial _{\lbrack \mu }\eta _{\nu ]},\eta _{\mu
}\eta _{\nu }\eta _{\rho }\eta _{\lambda },\eta _{\mu }\eta _{\nu }\eta
_{\rho }\partial _{\lbrack \alpha }\eta _{\beta ]},\eta _{\mu }\eta _{\nu
}\left( \partial _{\lbrack \alpha }\eta _{\beta ]}\right) \left( \partial
_{\lbrack \gamma }\eta _{\delta ]}\right) ,\right.  \notag \\
&&\left. \eta _{\mu }\left( \partial _{\lbrack \nu }\eta _{\rho ]}\right)
\left( \partial _{\lbrack \alpha }\eta _{\beta ]}\right) \left( \partial
_{\lbrack \gamma }\eta _{\delta ]}\right) ,\left( \partial _{\lbrack \mu
}\eta _{\nu ]}\right) \left( \partial _{\lbrack \rho }\eta _{\lambda
]}\right) \left( \partial _{\lbrack \alpha }\eta _{\beta ]}\right) \left(
\partial _{\lbrack \gamma }\eta _{\delta ]}\right) \right\} ,  \label{fo9}
\end{eqnarray}%
we must take $a_{4}^{\mathrm{h-A}}=0$ because in $D=11$ there is no Lorentz
scalar constructed as a linear combination of the elements present in (\ref%
{fo9}).

The next possible maximum value of the antighost number appearing in $a^{%
\mathrm{h-A}}$ is $I=3$
\begin{equation}
a^{\mathrm{h-A}}=a_{0}^{\mathrm{h-A}}+a_{1}^{\mathrm{h-A}}+a_{2}^{\mathrm{h-A%
}}+a_{3}^{\mathrm{h-A}},  \label{fo11}
\end{equation}%
where the terms from development (\ref{fo11}) satisfy equations (\ref{3.6})
and (\ref{3.4})--(\ref{3.5}) for $I=3$. According to the general results
established in the above, the general solution to (\ref{3.6}) for $I=3$
reads as
\begin{eqnarray}
a_{3}^{\mathrm{h-A}} &=&C^{\ast \sigma }\left[ f_{\sigma }^{\mu \nu \alpha
\beta }\eta _{\mu }\eta _{\nu }\partial _{\lbrack \alpha }\eta _{\beta
]}+f_{\sigma }^{\mu \alpha \beta \gamma \delta }\eta _{\mu }\left( \partial
_{\lbrack \alpha }\eta _{\beta ]}\right) \left( \partial _{\lbrack \gamma
}\eta _{\delta ]}\right) \right.  \notag \\
&&\left. +f_{\sigma }^{\mu \nu \rho }\eta _{\mu }\eta _{\nu }\eta _{\rho
}+f_{\sigma }^{\mu \nu \alpha \beta \gamma \delta }\left( \partial _{\lbrack
\mu }\eta _{\nu ]}\right) \left( \partial _{\lbrack \alpha }\eta _{\beta
]}\right) \left( \partial _{\lbrack \gamma }\eta _{\delta ]}\right) \right] .
\label{fo13}
\end{eqnarray}%
All the coefficients denoted by $f$ must be constant (neither derivative nor
depending on the spacetime coordinates). Recalling that we work in $D=11$
spacetime dimensions, we have no such constant Lorentz tensors, so $a_{3}^{%
\mathrm{h-A}}$ must vanish.

Assuming now that $a^{\mathrm{h-A}}$ stops at $I=2$, we have that the
solution to (\ref{fo2b}) reduces to
\begin{equation}
a^{\mathrm{h-A}}=a_{0}^{\mathrm{h-A}}+a_{1}^{\mathrm{h-A}}+a_{2}^{\mathrm{h-A%
}},  \label{fo14}
\end{equation}%
where the pieces present in (\ref{fo14}) are subject to equations (\ref{3.6}%
) and (\ref{3.4})--(\ref{3.5}) for $I=2$. The general solution to (\ref{3.6}%
) (up to $\gamma $-exact contributions) can be written in $D=11$ as
\begin{equation}
a_{2}^{\mathrm{h-A}}=C^{\ast \mu \nu }\left[ c_{1}\eta _{\mu }\eta _{\nu
}+c_{2}\left( \partial _{\lbrack \mu }\eta _{\rho ]}\right) \partial
_{\lbrack \nu }\eta _{\lambda ]}\sigma ^{\rho \lambda }\right] ,
\label{fo17}
\end{equation}%
where $c_{1}$ and $c_{2}$ are arbitrary, real constants. Using definitions (%
\ref{f8a})--(\ref{f8h}) we infer that
\begin{eqnarray}
\delta a_{2}^{\mathrm{h-A}} &=&\partial _{\rho }\left\{ -3A^{\ast \mu \nu
\rho }\left[ c_{1}\eta _{\mu }\eta _{\nu }+c_{2}\left( \partial _{\lbrack
\mu }\eta _{\alpha ]}\right) \partial _{\lbrack \nu }\eta _{\beta ]}\sigma
^{\alpha \beta }\right] \right\}  \notag \\
&&+\gamma \left[ 3c_{2}A^{\ast \mu \nu \rho }\left( \partial _{\lbrack \mu
}\eta _{\alpha ]}\right) \partial _{\lbrack \nu }^{\left. {}\right. }h_{\rho
]}^{\alpha }\right] -3c_{1}A^{\ast \mu \nu \rho }\eta _{\mu }\partial
_{\lbrack \nu }\eta _{\rho ]}.  \label{fo18}
\end{eqnarray}%
Comparing (\ref{3.4}) for $I=2$ with the right-hand side of (\ref{fo18}), we
observe that $a_{2}^{\mathrm{h-A}}$ of the form (\ref{fo17}) leads to a
consistent $a_{1}^{\mathrm{h-A}}$ if and only if
\begin{equation}
-3c_{1}A^{\ast \mu \nu \rho }\eta _{\mu }\partial _{\lbrack \nu }\eta _{\rho
]}=\gamma f_{1}+\partial _{\mu }t_{1}^{\mu }.  \label{parta2}
\end{equation}%
By taking the Euler-Lagrange derivative of both sides of (\ref{parta2}) with
respect to $A^{\ast \mu \nu \rho }$ and recalling that it commutes with $%
\gamma $, we arrive at
\begin{equation}
-3c_{1}\eta _{\mu }\partial _{\lbrack \nu }\eta _{\rho ]}=\gamma \left(
f_{0\mu \nu \rho }\right) ,  \label{parta21}
\end{equation}%
where
\begin{equation*}
f_{0\mu \nu \rho }=\frac{\delta ^{\mathrm{L}}f_{1}}{\delta A^{\ast \mu \nu
\rho }}.
\end{equation*}%
Since $\eta _{\mu }\partial _{\lbrack \nu }\eta _{\rho ]}$ is a nontrivial
object from $H\left( \gamma \right) $, it results that the left-hand side of
(\ref{parta21}) is $\gamma $-exact if and only if $c_{1}=0$. Therefore, the
only consistent solution to (\ref{3.6}) at antighost number two is
\begin{equation}
a_{2}^{\mathrm{h-A}}=c_{2}C^{\ast \mu \nu }\left( \partial _{\lbrack \mu
}\eta _{\rho ]}\right) \partial _{\lbrack \nu }\eta _{\lambda ]}\sigma
^{\rho \lambda }.  \label{a2}
\end{equation}%
Inserting (\ref{a2}) in (\ref{3.4}) for $I=2$, we derive
\begin{equation}
a_{1}^{\mathrm{h-A}}=-3c_{2}A^{\ast \mu \nu \rho }\left( \partial _{\lbrack
\mu }\eta _{\alpha ]}\right) \partial _{\lbrack \nu }^{\left. {}\right.
}h_{\rho ]}^{\alpha }+\bar{a}_{1}^{\mathrm{h-A}},  \label{fo21}
\end{equation}%
where $\bar{a}_{1}^{\mathrm{h-A}}$ represents the general solution to
equation (\ref{3.6}) for $I=1$. According to (\ref{3.10}) in pure ghost
number equal to one, it results that the most general form of $\bar{a}_{1}^{%
\mathrm{h-A}}$ as solution to (\ref{3.6}) for $I=1$ that might provide
effective cross-interactions can be written like
\begin{equation}
\bar{a}_{1}^{\mathrm{h-A}}=A^{\ast \mu \nu \rho }\left( M_{\mu \nu \rho
}^{\lambda }\eta _{\lambda }+M_{\mu \nu \rho }^{\alpha \beta }\partial
_{\lbrack \alpha }\eta _{\beta ]}\right) +h^{\ast \mu \nu }\left( \bar{M}%
_{\mu \nu }^{\lambda }\eta _{\lambda }+\bar{M}_{\mu \nu }^{\alpha \beta
}\partial _{\lbrack \alpha }\eta _{\beta ]}\right) ,  \label{fo24}
\end{equation}%
where the $M$-like functions may depend on linearized Riemann tensor, on the
field-strength of the Abelian three-form as well as on their spacetime
derivatives and satisfy obvious symmetry/antisymmetry properties. Using the
definitions of $\delta $ and $\gamma $, after some computations we obtain
that
\begin{equation}
\delta a_{1}^{\mathrm{h-A}}=\partial _{\mu }j_{1}^{\mu }+\gamma b_{0}+c_{0},
\label{fo25}
\end{equation}%
where we used the notations
\begin{eqnarray}
j_{1}^{\mu } &=&\frac{1}{2}F^{\mu \nu \rho \lambda }\left[ -c_{2}\left(
\partial _{\lbrack \nu }\eta _{\alpha ]}\right) \partial _{\lbrack \rho
}^{\left. {}\right. }h_{\lambda ]}^{\alpha }+\frac{1}{3}\left( M_{\nu \rho
\lambda }^{\alpha }\eta _{\alpha }+M_{\nu \rho \lambda }^{\alpha \beta
}\partial _{\lbrack \alpha }\eta _{\beta ]}\right) \right]  \notag \\
&&-2\left( \partial _{\nu }\phi ^{\mu \alpha \nu \beta }\right) \left( \bar{M%
}_{\alpha \beta }^{\lambda }\eta _{\lambda }+\bar{M}_{\alpha \beta }^{\rho
\lambda }\partial _{\lbrack \rho }\eta _{\lambda ]}\right)  \notag \\
&&+2\phi ^{\mu \alpha \nu \beta }\partial _{\nu }\left( \bar{M}_{\alpha
\beta }^{\lambda }\eta _{\lambda }+\bar{M}_{\alpha \beta }^{\rho \lambda
}\partial _{\lbrack \rho }\eta _{\lambda ]}\right) ,  \label{fo26}
\end{eqnarray}%
\begin{eqnarray}
b_{0} &=&F^{\mu \nu \rho \lambda }\left[ -\frac{c_{2}}{8}\left( \partial
_{\lbrack \mu }h_{\nu ]\alpha }\right) \partial _{\lbrack \rho }^{\left.
{}\right. }h_{\lambda ]}^{\alpha }+\frac{1}{6}\left( M_{\mu \nu \rho
}^{\alpha \beta }\partial _{\lbrack \alpha }h_{\beta ]\lambda }+\frac{1}{2}%
M_{\mu \nu \rho }^{\alpha }h_{\alpha \lambda }\right) \right]  \notag \\
&&+2\phi ^{\mu \alpha \nu \beta }\left[ \frac{1}{2}h_{\beta \lambda
}\partial _{\lbrack \mu }^{\left. {}\right. }\bar{M}_{\alpha ]\nu }^{\lambda
}-\bar{M}_{\mu \nu }^{\lambda }\overset{(1)}{\Gamma }_{\lambda \alpha \beta
}-\bar{M}_{\mu \nu }^{\rho \lambda }\partial _{\lbrack \rho }\overset{(1)}{%
\Gamma }_{\lambda ]\alpha \beta }\right.  \notag \\
&&\left. +\left( \partial _{\lbrack \mu }\bar{M}_{\alpha ]\nu }^{\rho
\lambda }\right) \left( \partial _{\lbrack \rho }h_{\lambda ]\beta }\right) %
\right] ,  \label{fo27}
\end{eqnarray}%
\begin{eqnarray}
c_{0} &=&-\frac{1}{4!}F^{\mu \nu \rho \lambda }\left[ \eta _{\alpha
}\partial _{\lbrack \mu }^{\left. {}\right. }M_{\nu \rho \lambda ]}^{\alpha
}+\left( \partial _{\lbrack \mu }^{\left. {}\right. }M_{\nu \rho \lambda
]}^{\alpha \beta }+2\delta _{\lambda }^{\beta }M_{\mu \nu \rho }^{\alpha
}\right) \partial _{\lbrack \alpha }\eta _{\beta ]}\right]  \notag \\
&&+\frac{1}{2}\phi ^{\mu \alpha \nu \beta }\left[ \left( \partial _{\lbrack
\mu }^{\left. {}\right. }\bar{M}_{\alpha ][\nu ,\beta ]}^{\rho \lambda
}+2\delta _{\beta }^{\rho }\partial _{\lbrack \mu }^{\left. {}\right. }\bar{M%
}_{\alpha ]\nu }^{\lambda }\right) \partial _{\lbrack \rho }\eta _{\lambda
]}+\eta _{\lambda }\partial _{\lbrack \mu }^{\left. {}\right. }\bar{M}%
_{\alpha ][\nu ,\beta ]}^{\rho \lambda }\right] ,  \label{fo28}
\end{eqnarray}%
\begin{equation}
\phi ^{\mu \alpha \nu \beta }=\frac{1}{2}\left( h^{\alpha \lbrack \nu
}\sigma ^{\beta ]\mu }-h^{\mu \lbrack \nu }\sigma ^{\beta ]\alpha }+h\sigma
^{\mu \lbrack \nu }\sigma ^{\beta ]\alpha }\right) ,  \label{fo30}
\end{equation}%
\begin{equation}
\overset{(1)}{\Gamma }_{\lambda \alpha \beta }=\frac{1}{2}\left( \partial
_{\alpha }h_{\beta \lambda }+\partial _{\beta }h_{\alpha \lambda }-\partial
_{\lambda }h_{\alpha \beta }\right) .  \label{wgt}
\end{equation}%
According to (\ref{3.5}) for $I=1$, (\ref{fo27}) gives (up to a global
factor) some of the pieces from the interacting Lagrangian at order one in
the coupling constant. The hypothesis on the maximum number of derivatives
in the interacting Lagrangian being equal to two induces further
restrictions on the type-$M$ functions, as it will be seen bellow. The first
term from (\ref{fo27}) outputs an interacting vertex with three derivatives,
which disagrees with this hypothesis. Therefore, we must annihilate the
corresponding constant, $c_{2}=0$. In order to provide cross-couplings, the
functions $\bar{M}_{\alpha \beta }^{\rho \lambda }$ and $\bar{M}_{\mu \nu
}^{\lambda }$ must effectively depend on the field-strength of the Abelian
three-form. Consequently, the last two terms on the right-hand side of (\ref%
{fo27}) will produce terms with at least three derivatives in the
interacting Lagrangian, so we must discard them by setting $\bar{M}_{\alpha
\beta }^{\rho \lambda }=0$. If we represent the functions $\bar{M}_{\mu \nu
}^{\lambda }$ as
\begin{equation*}
\bar{M}_{\mu \nu }^{\lambda }=f_{\mu \nu }^{\lambda \alpha \beta \gamma
\delta }F_{\alpha \beta \gamma \delta },
\end{equation*}%
where $f_{\mu \nu }^{\lambda \alpha \beta \gamma \delta }$ are nonderivative
Lorentz constants, we conclude that we have no such constant tensors in $%
D=11 $, so we must take $\bar{M}_{\mu \nu }^{\lambda }=0$. The pieces from (%
\ref{fo27}) proportional with $M_{\mu \nu \rho }^{\alpha \beta }$ satisfy
the assumption on the derivative order if and only if these functions are
nonderivative Lorentz constants. Since in $D=11$ there are no such constant
tensors, we conclude that we must take $M_{\mu \nu \rho }^{\alpha \beta }=0$%
. Finally, the functions $M_{\mu \nu \rho }^{\alpha }$ produce terms in the
interacting Lagrangian that comply with the hypothesis on the maximum number
of derivatives if and only if they are linear in the undifferentiated
field-strength of the Abelian three-form. Due to the spacetime dimension,
there is just one possibility left, namely
\begin{equation}
M_{\mu \nu \rho }^{\alpha }=k\sigma ^{\alpha \beta }F_{\mu \nu \rho \beta },
\label{r4}
\end{equation}%
where $k$ is an arbitrary, real constant.

Inserting the above results in (\ref{a2}) and (\ref{fo24}), we infer
\begin{eqnarray}
a_{2}^{\mathrm{h-A}} &=&0,  \label{fo31} \\
a_{1}^{\mathrm{h-A}} &=&kA^{*\mu \nu \rho }F_{\mu \nu \rho \lambda }\eta
^{\lambda }.  \label{fo32}
\end{eqnarray}
Applying now the Koszul-Tate operator $\delta $ on (\ref{fo32}), we
determine the interacting Lagrangian at order one in the coupling constant
as
\begin{equation}
a_{0}^{\mathrm{h-A}}=-\frac{k}{12}F^{\mu \nu \rho \lambda }\left( F_{\mu \nu
\rho \sigma }h_{\lambda }^{\sigma }-\frac{1}{8}F_{\mu \nu \rho \lambda
}h\right) .  \label{fo34}
\end{equation}
By assembling the previous results we can state that the general solution to
(\ref{fo2b}) in $D=11$ reads as
\begin{equation}
a^{\mathrm{h-A}}=kA^{*\mu \nu \rho }F_{\mu \nu \rho \lambda }\eta ^{\lambda
}-\frac{k}{12}F^{\mu \nu \rho \lambda }\left( F_{\mu \nu \rho \sigma
}h_{\lambda }^{\sigma }-\frac{1}{8}F_{\mu \nu \rho \lambda }h\right) .
\label{fo37}
\end{equation}
We can still remove from (\ref{fo37}) certain trivial, $s$-exact modulo $d$
terms. Indeed, we have that
\begin{eqnarray}
a^{\mathrm{h-A}} &=&\partial _{\mu }\left[ -\frac{k}{4}F^{\mu \nu \rho
\lambda }A_{\nu \rho \sigma }h_{\lambda }^{\sigma }+3kA^{*\mu \nu \rho
}\left( A_{\nu \rho \lambda }\eta ^{\lambda }+C_{\nu \lambda }h_{\rho
}^{\lambda }\right) \right.  \notag \\
&&\left. +kC^{*\mu \nu }\left( C^{\rho }h_{\nu \rho }-2C_{\nu \rho }\eta
^{\rho }\right) +kC^{*}C^{\nu }\eta _{\nu }\right]  \notag \\
&&+s\left[ -\frac{3k}{2}A^{*\mu \nu \rho }A_{\mu \nu \lambda }h_{\rho
}^{\lambda }-kC^{*\mu \nu }\left( A_{\mu \nu \rho }\eta ^{\rho }+C_{\mu
\lambda }h_{\nu }^{\lambda }\right) \right.  \notag \\
&&\left. +kC^{*\mu }\left( C_{\mu \nu }\eta ^{\nu }-\frac{1}{2}C^{\nu
}h_{\mu \nu }\right) -kC^{*}C^{\mu }\eta _{\mu }\right]  \notag \\
&&+\frac{k}{12}F^{\mu \nu \rho \lambda }\left( 3\partial _{\mu }\left(
A_{\nu \rho \sigma }h_{\lambda }^{\sigma }\right) -F_{\mu \nu \rho \sigma
}h_{\lambda }^{\sigma }+\frac{1}{8}F_{\mu \nu \rho \lambda }h\right)  \notag
\\
&&-\frac{3k}{2}A^{*\mu \nu \rho }\left( \frac{2}{3}\eta ^{\lambda }\partial
_{\lambda }A_{\mu \nu \rho }+A_{\mu \nu }^{\;\;\;\lambda }\partial _{[\rho
}\eta _{\lambda ]}-h_{\rho \lambda }\partial ^{\lambda }C_{\mu \nu }-C_{\mu
\lambda }\partial _{[\nu }^{\left. {}\right. }h_{\rho ]}^{\lambda }\right)
\notag \\
&&-kC^{*\mu \nu }\left[ \left( \partial _{\rho }C_{\mu \nu }\right) \eta
^{\rho }+C_{\mu }^{\;\;\rho }\partial _{[\nu }\eta _{\rho ]}+h_{\nu \rho
}\partial ^{\rho }C_{\mu }+\frac{1}{2}C^{\rho }\partial _{[\mu }h_{\nu ]\rho
}\right]  \notag \\
&&-\frac{k}{2}C^{*\mu }\left( 2\eta _{\nu }\partial ^{\nu }C_{\mu }+C^{\nu
}\partial _{[\mu }\eta _{\nu ]}-h_{\mu \nu }\partial ^{\nu }C\right)
-kC^{*}\left( \partial ^{\mu }C\right) \eta _{\mu }.  \label{fo38}
\end{eqnarray}
Since $S_{1}$ is unique up to $s$-exact modulo $d$ terms (see subsection \ref%
{stand}), we can remove such terms and work, instead of (\ref{fo37}), with
\begin{eqnarray}
&&a^{\mathrm{h-A}}=\frac{k}{12}F^{\mu \nu \rho \lambda }\left[ 3\partial
_{\mu }\left( A_{\nu \rho \sigma }h_{\lambda }^{\sigma }\right) -F_{\mu \nu
\rho \sigma }h_{\lambda }^{\sigma }+\frac{1}{8}F_{\mu \nu \rho \lambda }h%
\right]  \notag \\
&&-\frac{3k}{2}A^{*\mu \nu \rho }\left( \frac{2}{3}\eta ^{\lambda }\partial
_{\lambda }A_{\mu \nu \rho }+A_{\mu \nu }^{\;\;\;\lambda }\partial _{[\rho
}\eta _{\lambda ]}-h_{\rho \lambda }\partial ^{\lambda }C_{\mu \nu }-C_{\mu
\lambda }\partial _{[\nu }^{\left. {}\right. }h_{\rho ]}^{\lambda }\right)
\notag \\
&&-kC^{*\mu \nu }\left[ \left( \partial _{\rho }C_{\mu \nu }\right) \eta
^{\rho }+C_{\mu }^{\;\;\rho }\partial _{[\nu }\eta _{\rho ]}+h_{\nu \rho
}\partial ^{\rho }C_{\mu }+\frac{1}{2}C^{\rho }\partial _{[\mu }h_{\nu ]\rho
}\right]  \notag \\
&&-\frac{k}{2}C^{*\mu }\left( 2\eta _{\nu }\partial ^{\nu }C_{\mu }+C^{\nu
}\partial _{[\mu }\eta _{\nu ]}-h_{\mu \nu }\partial ^{\nu }C\right)
-kC^{*}\left( \partial ^{\mu }C\right) \eta _{\mu }.  \label{fo39}
\end{eqnarray}

The above results can be summarized by the conclusion that the `interacting'
part of the first-order deformation of the solution to the master equation
can be written as
\begin{equation}
S_{1}^{\prime \mathrm{h-A}}=\int d^{11}x\left( a^{\mathrm{h-A}}+a_{0}^{%
\mathrm{A}}\right) ,  \label{frstord}
\end{equation}
where $a^{\mathrm{h-A}}$ is given in (\ref{fo39}) and $a_{0}^{\mathrm{A}}$
is expressed by (\ref{fo5}).

\subsection{Second-order deformation\label{secondord}}

Until now we have seen that the first-order deformation can be written like
the sum between the Pauli-Fierz component $S_{1}^{\mathrm{h}}$ (given in
detail in Ref.~\cite{multi}) and the 'interacting' part $S_{1}^{\prime
\mathrm{h-A}}$, expressed by (\ref{frstord}).

In this section we investigate the consistency of the first-order
deformation, described by equation (\ref{2.6}). Along the same line as
before, we can write the second-order deformation like the sum between the
Pauli-Fierz contribution and the interacting part
\begin{equation}
S_{2}^{\mathrm{h,A}}=S_{2}^{\mathrm{h}}+S_{2}^{\mathrm{h-A}}.  \label{4.2}
\end{equation}%
The piece $S_{2}^{\mathrm{h}}$ can be deduced from Ref.~\cite{multi}, while $%
S_{2}^{\mathrm{h-A}}$ is subject to the equation
\begin{equation}
\frac{1}{2}\left( S_{1},S_{1}\right) ^{\mathrm{h-A}}+sS_{2}^{\mathrm{h-A}}=0,
\label{4.3}
\end{equation}%
where
\begin{equation}
\left( S_{1},S_{1}\right) ^{\mathrm{h-A}}=\left( S_{1}^{\prime \mathrm{h-A}%
},S_{1}^{\prime \mathrm{h-A}}\right) +2\left( S_{1}^{\mathrm{h}%
},S_{1}^{\prime \mathrm{h-A}}\right) .  \label{4.4}
\end{equation}%
If we denote by $\Delta ^{\mathrm{h-A}}$ and $b^{\mathrm{h-A}}$ the
nonintegrated densities of the functionals $\left( S_{1},S_{1}\right) ^{%
\mathrm{h-A}}$ and respectively $S_{2}^{\mathrm{h-A}}$, then the local form
of (\ref{4.3}) becomes
\begin{equation}
\Delta ^{\mathrm{h-A}}=-2sb^{\mathrm{h-A}}+\partial _{\mu }n^{\mu },
\label{4.5}
\end{equation}%
with
\begin{equation}
\mathrm{gh}\left( \Delta ^{\mathrm{h-A}}\right) =1,\qquad \mathrm{gh}\left(
b^{\mathrm{h-A}}\right) =0,\qquad \mathrm{gh}\left( n^{\mu }\right) =1,
\label{45a}
\end{equation}%
for some local currents $n^{\mu }$. Direct computation shows that $\Delta ^{%
\mathrm{h-A}}$ decomposes like
\begin{equation}
\Delta ^{\mathrm{h-A}}=\sum_{I=0}^{4}\Delta _{I}^{\mathrm{h-A}},\qquad
\mathrm{agh}\left( \Delta _{I}^{\mathrm{h-A}}\right) =I,\qquad I=\overline{%
0,4},  \label{4.6}
\end{equation}%
with
\begin{equation}
\Delta _{4}^{\mathrm{h-A}}=\gamma \left[ -k^{2}C^{\ast }h_{\mu \nu }\eta
^{\mu }\partial ^{\nu }C+k\left( k+1\right) C^{\ast }C^{\mu }\left( \partial
_{\lbrack \mu }\eta _{\nu ]}\right) \eta ^{\nu }\right] +\partial _{\mu
}\tau _{4}^{\mu },  \label{4.7}
\end{equation}%
\begin{eqnarray}
\Delta _{3}^{\mathrm{h-A}} &=&\delta \left[ -k^{2}C^{\ast }h_{\mu \nu }\eta
^{\mu }\partial ^{\nu }C+k\left( k+1\right) C^{\ast }C^{\mu }\left( \partial
_{\lbrack \mu }\eta _{\nu ]}\right) \eta ^{\nu }\right]  \notag \\
&&+\gamma \left\{ \frac{k}{2}C^{\ast \mu }\left[ \frac{1-2k}{2}\left(
\partial ^{\nu }C\right) h_{\mu \rho }h_{\nu }^{\rho }-kC^{\nu }h_{\mu
}^{\rho }\partial _{\lbrack \nu }\eta _{\rho ]}+kC^{\nu }\eta ^{\rho
}\partial _{\lbrack \mu }h_{\nu ]\rho }\right. \right.  \notag \\
&&-\frac{1}{2}C^{\nu }\left( h_{\mu }^{\rho }\partial _{\lbrack \nu }\eta
_{\rho ]}+h_{\nu }^{\rho }\partial _{\lbrack \mu }\eta _{\rho ]}\right)
+\left( k+1\right) C^{\nu }\eta ^{\rho }\left( \partial _{\lbrack \mu
}h_{\rho ]\nu }+\partial _{\lbrack \nu }h_{\rho ]\mu }\right)  \notag \\
&&\left. \left. +2k\left( \partial _{\nu }C_{\mu }\right) h^{\nu \rho }\eta
_{\rho }-2\left( k+1\right) C_{\mu \nu }\left( \partial ^{\lbrack \nu }\eta
^{\rho ]}\right) \eta _{\rho }\right] \right\} +\partial _{\mu }\tau
_{3}^{\mu },  \label{4.8}
\end{eqnarray}%
\begin{eqnarray}
\Delta _{2}^{\mathrm{h-A}} &=&\delta \left\{ \frac{k}{2}C^{\ast \mu }\left[
\frac{1-2k}{2}\left( \partial ^{\nu }C\right) h_{\mu \rho }h_{\nu }^{\rho
}-kC^{\nu }h_{\mu }^{\rho }\partial _{\lbrack \nu }\eta _{\rho ]}+kC^{\nu
}\eta ^{\rho }\partial _{\lbrack \mu }h_{\nu ]\rho }\right. \right.  \notag
\\
&&-\frac{1}{2}C^{\nu }\left( h_{\mu }^{\rho }\partial _{\lbrack \nu }\eta
_{\rho ]}+h_{\nu }^{\rho }\partial _{\lbrack \mu }\eta _{\rho ]}\right)
+\left( k+1\right) C^{\nu }\eta ^{\rho }\left( \partial _{\lbrack \mu
}h_{\rho ]\nu }+\partial _{\lbrack \nu }h_{\rho ]\mu }\right)  \notag \\
&&\left. \left. +2k\left( \partial _{\nu }C_{\mu }\right) h^{\nu \rho }\eta
_{\rho }-2\left( k+1\right) C_{\mu \nu }\left( \partial ^{\lbrack \nu }\eta
^{\rho ]}\right) \eta _{\rho }\right] \right\}  \notag \\
&&+\gamma \left\{ kC^{\ast \mu \nu }\left[ -\frac{1}{2}C^{\rho }\partial
_{\mu }\left( h_{\nu \lambda }h_{\rho }^{\lambda }\right) +kC^{\rho }h_{\nu
}^{\lambda }\partial _{\lbrack \mu }h_{\lambda ]\rho }\right. \right.  \notag
\\
&&+\left( k-\frac{1}{2}\right) \left( \partial _{\rho }C_{\mu }\right)
h_{\nu \lambda }h^{\rho \lambda }+\eta ^{\rho }\left( \partial _{\lbrack \mu
}^{\left. {}\right. }h_{\rho ]}^{\lambda }+\partial _{\lbrack \xi }h_{\rho
]\mu }\sigma ^{\xi \lambda }\right) C_{\nu \lambda }  \notag \\
&&+\frac{1}{2}C_{\nu \rho }\left( h^{\rho \lambda }\partial _{\lbrack \mu
}\eta _{\lambda ]}+h_{\mu \lambda }\partial ^{\lbrack \rho }\eta ^{\lambda
]}\right) -kC_{\mu \rho }\left( 2\eta ^{\lambda }\partial _{\lbrack \lambda
}^{\left. {}\right. }h_{\nu ]}^{\rho }+h_{\nu \lambda }\partial ^{\lbrack
\rho }\eta ^{\lambda ]}\right)  \notag \\
&&\left. \left. +k\left( \partial _{\rho }C_{\mu \nu }\right) \eta _{\lambda
}h^{\rho \lambda }-\left( k+1\right) A_{\mu \nu \rho }\eta _{\lambda
}\partial ^{\lbrack \rho }\eta ^{\lambda ]}\right] \right\} +\partial _{\mu
}\tau _{2}^{\mu },  \label{4.9}
\end{eqnarray}%
\begin{eqnarray}
\Delta _{1}^{\mathrm{h-A}} &=&\delta \left\{ kC^{\ast \mu \nu }\left[ -\frac{%
1}{2}C^{\rho }\partial _{\mu }\left( h_{\nu \lambda }h_{\rho }^{\lambda
}\right) +kC^{\rho }h_{\nu }^{\lambda }\partial _{\lbrack \mu }h_{\lambda
]\rho }+k\left( \partial _{\rho }C_{\mu \nu }\right) \eta _{\lambda }h^{\rho
\lambda }\right. \right.  \notag \\
&&+\left( k-\frac{1}{2}\right) \left( \partial _{\rho }C_{\mu }\right)
h_{\nu \lambda }h^{\rho \lambda }+\eta ^{\rho }\left( \partial _{\lbrack \mu
}^{\left. {}\right. }h_{\rho ]}^{\lambda }+\partial _{\lbrack \xi }h_{\rho
]\mu }\sigma ^{\xi \lambda }\right) C_{\nu \lambda }  \notag \\
&&+\frac{1}{2}C_{\nu \rho }\left( h^{\rho \lambda }\partial _{\lbrack \mu
}\eta _{\lambda ]}+h_{\mu \lambda }\partial ^{\lbrack \rho }\eta ^{\lambda
]}\right) -kC_{\mu \rho }\left( 2\eta ^{\lambda }\partial _{\lbrack \lambda
}^{\left. {}\right. }h_{\nu ]}^{\rho }+h_{\nu \lambda }\partial ^{\lbrack
\rho }\eta ^{\lambda ]}\right)  \notag \\
&&\left. \left. -\left( k+1\right) A_{\mu \nu \rho }\eta _{\lambda }\partial
^{\lbrack \rho }\eta ^{\lambda ]}\right] \right\} +\gamma \left\{ \frac{3}{2}%
kA^{\ast \mu \nu \rho }\left[ C_{\rho \xi }\partial _{\mu }\left( h_{\nu
\lambda }h^{\lambda \xi }\right) \right. \right.  \notag \\
&&-2kC_{\mu \lambda }h_{\nu }^{\xi }\partial _{\lbrack \xi }^{\left.
{}\right. }h_{\rho ]}^{\lambda }+\frac{1-2k}{2}h_{\rho \xi }h^{\lambda \xi
}\partial _{\lambda }C_{\mu \nu }  \notag \\
&&-\frac{1}{2}A_{\mu \nu \lambda }\left( h^{\lambda \xi }\partial _{\lbrack
\rho }\eta _{\xi ]}+h_{\rho \xi }\partial ^{\lbrack \lambda }\eta ^{\xi
]}\right) +A_{\mu \nu \lambda }\eta _{\xi }\left( 2\partial _{\left.
{}\right. }^{[\lambda }h_{\rho }^{\xi ]}-\sigma ^{\lambda \pi }\partial
_{\lbrack \pi }h_{\rho ]\xi }\right)  \notag \\
&&\left. \left. +kA_{\mu \nu \lambda }\left( 2\eta ^{\xi }\partial _{\lbrack
\rho }^{\left. {}\right. }h_{\xi ]}^{\lambda }-h_{\rho \xi }\partial
^{\lbrack \lambda }\eta ^{\xi ]}\right) +\frac{2}{3}k\left( \partial
_{\lambda }A_{\mu \nu \rho }\right) h^{\lambda \xi }\eta _{\xi }\right]
\right\}  \notag \\
&&-k\left( k+1\right) A^{\ast \mu \nu \rho }F_{\mu \nu \rho \lambda }\left(
\partial ^{\lbrack \lambda }\eta ^{\xi ]}\right) \eta _{\xi }+\partial _{\mu
}\tau _{1}^{\mu },  \label{4.10}
\end{eqnarray}%
and
\begin{eqnarray}
\Delta _{0}^{\mathrm{h-A}} &=&\delta \left\{ \frac{3}{2}kA^{\ast \mu \nu
\rho }\left[ C_{\rho \xi }\partial _{\mu }\left( h_{\nu \lambda }h^{\lambda
\xi }\right) -2kC_{\mu \lambda }h_{\nu }^{\xi }\partial _{\lbrack \xi
}^{\left. {}\right. }h_{\rho ]}^{\lambda }\right. \right.  \notag \\
&&+\frac{1-2k}{2}h_{\rho \xi }h^{\lambda \xi }\partial _{\lambda }C_{\mu \nu
}-\frac{1}{2}A_{\mu \nu \lambda }\left( h^{\lambda \xi }\partial _{\lbrack
\rho }\eta _{\xi ]}+h_{\rho \xi }\partial ^{\lbrack \lambda }\eta ^{\xi
]}\right)  \notag \\
&&+A_{\mu \nu \lambda }\eta _{\xi }\left( 2\partial _{\left. {}\right.
}^{[\lambda }h_{\rho }^{\xi ]}-\sigma ^{\lambda \pi }\partial _{\lbrack \pi
}h_{\rho ]\xi }\right) +kA_{\mu \nu \lambda }\left( 2\eta ^{\xi }\partial
_{\lbrack \rho }^{\left. {}\right. }h_{\xi ]}^{\lambda }-h_{\rho \xi
}\partial ^{\lbrack \lambda }\eta ^{\xi ]}\right)  \notag \\
&&\left. \left. +\frac{2k}{3}\left( \partial _{\lambda }A_{\mu \nu \rho
}\right) h^{\lambda \xi }\eta _{\xi }\right] \right\} +\gamma \left\{ \frac{k%
}{4!}F^{\mu \nu \rho \lambda }F_{\mu \nu \xi \pi }\left( -3h_{\rho }^{\xi
}h_{\lambda }^{\pi }-\delta _{\rho }^{\xi }h_{\lambda \sigma }h^{\pi \sigma
}\right. \right.  \notag \\
&&\left. +\frac{1-k}{8}\delta _{\rho }^{\xi }\delta _{\lambda }^{\pi
}h_{\alpha \beta }h^{\alpha \beta }+\frac{k}{8}\delta _{\rho }^{\xi }\delta
_{\lambda }^{\pi }h^{2}\right) +\frac{k}{2}F^{\mu \nu \rho \lambda }\left[
\frac{1}{4}A_{\xi \rho \lambda }\partial _{\mu }\left( h_{\nu \pi
}h^{\lambda \pi }\right) \right.  \notag \\
&&-\frac{1}{4}h_{\mu \pi }h^{\lambda \pi }\partial _{\nu }A_{\xi \rho
\lambda }-\frac{k}{3}h_{\lambda \pi }h^{\xi \pi }\partial _{\xi }A_{\mu \nu
\rho }-kA_{\mu \nu \xi }h_{\rho }^{\pi }\partial _{\lbrack \pi }^{\left.
{}\right. }h_{\lambda ]}^{\xi }  \notag \\
&&-\frac{k}{6}h_{\mu }^{\xi }h\partial _{\xi }A_{\nu \rho \lambda }+\frac{k}{%
4}A_{\mu \nu \xi }h\partial _{\lbrack \rho }^{\left. {}\right. }h_{\lambda
]}^{\xi }  \notag \\
&&\left. +\frac{k}{2}A_{\mu \xi \pi }\partial _{\nu }\left( h_{\rho }^{\xi
}h_{\lambda }^{\pi }\right) +\frac{k}{2}h_{\rho }^{\xi }h_{\lambda }^{\pi
}\partial _{\xi }A_{\pi \mu \nu }\right]  \notag \\
&&-\frac{k^{2}}{8}\partial _{\xi }\left( h_{[\mu }^{\pi }A_{\nu \rho ]\pi
}^{\left. {}\right. }\right) \left[ \partial ^{\rho }\left( h_{\tau }^{[\xi
}A_{\left. {}\right. }^{\mu \nu ]\tau }\right) -\frac{1}{3}\partial ^{\xi
}\left( h_{\tau }^{[\mu }A_{\left. {}\right. }^{\nu \rho ]\tau }\right) %
\right]  \notag \\
&&+kq\varepsilon ^{\mu _{1}\ldots \mu _{11}}\left( hA_{\mu _{1}\mu _{2}\mu
_{3}}F_{\mu _{4}\ldots \mu _{7}}-8h_{\mu _{1}}^{\xi }A_{\mu _{2}\mu _{3}\mu
_{4}}F_{\mu _{5}\ldots \mu _{7}\xi }\right.  \notag \\
&&\left. \left. +6h_{\mu _{1}}^{\xi }A_{\xi \mu _{2}\mu _{3}}F_{\mu
_{4}\ldots \mu _{7}}\right) F_{\mu _{8}\ldots \mu _{11}}\right\} -\frac{1}{3!%
}k\left( k+1\right) F^{\mu \nu \rho \lambda }F_{\mu \nu \rho \xi }\times
\notag \\
&&\times \left[ \eta ^{\pi }\left( \frac{1}{8}\delta _{\lambda }^{\xi
}\partial _{\lbrack \sigma }^{\left. {}\right. }h_{\pi ]}^{\sigma }-\partial
_{\lbrack \lambda }h_{\pi ]}^{\xi }\right) +\frac{1}{2}h_{\lambda \pi
}\partial ^{\lbrack \xi }\eta ^{\pi ]}\right] +\partial _{\mu }\tau
_{0}^{\mu }.  \label{4.11}
\end{eqnarray}%
Because $\left( S_{1},S_{1}\right) ^{\mathrm{h-A}}$ contains terms of
maximum antighost number equal to four, we can assume (without loss of
generality) that $b^{\mathrm{h-A}}$ stops at antighost number five%
\begin{eqnarray}
b^{\mathrm{h-A}} &=&\sum_{I=0}^{5}b_{I}^{\mathrm{h-A}},\qquad \mathrm{agh}%
\left( b_{I}^{\mathrm{h-A}}\right) =I,\qquad I=\overline{0,5},  \label{4.12}
\\
n^{\mu } &=&\sum_{I=0}^{5}n_{I}^{\mu },\qquad \mathrm{agh}\left( n_{I}^{\mu
}\right) =I,\qquad I=\overline{0,5}.  \label{4.13}
\end{eqnarray}%
By projecting equation (\ref{4.5}) on the various (decreasing) values of the
antighost number, we infer the following tower of equations
\begin{eqnarray}
\gamma b_{5}^{\mathrm{h-A}} &=&\partial _{\mu }\left( \frac{1}{2}n_{5}^{\mu
}\right) ,  \label{4.14} \\
\Delta _{I}^{\mathrm{h-A}} &=&-2\left( \delta b_{I+1}^{\mathrm{h-A}}+\gamma
b_{I}^{\mathrm{h-A}}\right) +\partial _{\mu }n_{I}^{\mu },\qquad I=\overline{%
0,4}.  \label{4.15}
\end{eqnarray}%
Equation (\ref{4.14}) can always be replaced with
\begin{equation}
\gamma b_{5}^{\mathrm{h-A}}=0.  \label{a74xz}
\end{equation}%
If we compare (\ref{4.7}) with (\ref{4.15}) for $I=4$, then we find that $%
b_{5}^{\mathrm{h-A}}$ is restricted to fulfill the equation
\begin{equation}
\delta b_{5}^{\mathrm{h-A}}+\gamma \tilde{b}_{4}^{\mathrm{h-A}}=\partial
_{\mu }\tilde{n}_{4}^{\mu },  \label{wa1}
\end{equation}%
where
\begin{equation}
b_{4}^{\mathrm{h-A}}=-\frac{1}{2}\left[ -k^{2}C^{\ast }h_{\mu \nu }\eta
^{\mu }\partial ^{\nu }C+k\left( k+1\right) C^{\ast }C^{\mu }\left( \partial
_{\lbrack \mu }\eta _{\nu ]}\right) \eta ^{\nu }\right] +\tilde{b}_{4}^{%
\mathrm{h-A}}.  \label{wa2}
\end{equation}%
By (\ref{3.10}) we get that the solution to (\ref{a74xz}) reads as
\begin{equation}
b_{5}^{\mathrm{h-A}}=\bar{\alpha}_{5}(\left[ F_{\mu \nu \rho \lambda }\right]
,\left[ K_{\mu \nu \alpha \beta }\right] ,\left[ \chi _{\Delta }^{\ast }%
\right] )\omega ^{5}\left( C,\eta _{\mu },\partial _{\lbrack \mu }\eta _{\nu
]}\right) .  \label{b5interm}
\end{equation}%
Substituting the above form of $b_{5}^{\mathrm{h-A}}$ into (\ref{wa1}), we
infer that a necessary condition for (\ref{wa1}) to possess solutions is
that $\bar{\alpha}_{5}$ belongs to $H_{5}\left( \delta |d\right) $. Since
for the model under consideration we know that $H_{5}\left( \delta |d\right)
=0$ and $H_{5}^{\mathrm{inv}}\left( \delta |d\right) =0$, it follows that we
can take
\begin{equation}
b_{5}^{\mathrm{h-A}}=0,  \label{wa3}
\end{equation}%
such that equation (\ref{wa1}) reduces to $\gamma \tilde{b}_{4}^{\mathrm{h-A}%
}=\partial _{\mu }\tilde{n}_{4}^{\mu }$. The last equation can always be
replaced (as it stands in a strictly positive value of the antighost number)
with $\gamma \tilde{b}_{4}^{\mathrm{h-A}}=0$. The last equation was
investigated in the previous subsection and was shown to possess only the
trivial solution
\begin{equation}
\tilde{b}_{4}^{\mathrm{h-A}}=0.  \label{wa4}
\end{equation}%
Due to (\ref{wa3}) and (\ref{wa4}), we observe that relations (\ref{4.7})--(%
\ref{4.9}) agree with equation (\ref{4.15}) for $I=4$, $I=3$ and $I=2$
respectively. On the contrary, $\Delta _{1}^{\mathrm{h-A}}$ given in (\ref%
{4.10}) cannot be written like in (\ref{4.15}) for $I=1$ unless
\begin{equation}
\chi =-k\left( k+1\right) A^{\ast \mu \nu \rho }F_{\mu \nu \rho \lambda
}\left( \partial ^{\lbrack \lambda }\eta ^{\xi ]}\right) \eta _{\xi },
\label{4.16}
\end{equation}%
can be expressed like
\begin{equation}
\chi =\delta \varphi +\gamma \omega +\partial _{\mu }l^{\mu .}.  \label{4.17}
\end{equation}%
Assume that (\ref{4.17}) holds. Then, by acting with $\delta $ on it from
the left, we infer that
\begin{equation}
\delta \chi =\gamma \left( -\delta \omega \right) +\partial _{\mu }\left(
\delta l^{\mu }\right) .  \label{4.18}
\end{equation}%
On the other hand, using the concrete expression of $\chi $, we have that
\begin{equation}
\delta \chi =k\left( k+1\right) \left\{ \gamma \left[ -T_{\lambda }^{\pi
}\left( \eta _{\xi }\partial _{\left. {}\right. }^{[\lambda }h_{\pi }^{\xi
]}+\frac{1}{2}h_{\pi \xi }\partial ^{\lbrack \lambda }\eta ^{\xi ]}\right) %
\right] +\partial _{\mu }\left( T_{\tau }^{\mu }\eta _{\xi }\partial
^{\lbrack \tau }\eta ^{\xi ]}\right) \right\} ,  \label{4.19}
\end{equation}%
where
\begin{equation}
T^{\alpha \beta }=\frac{1}{3!}F^{\mu \nu \rho \alpha }F_{\mu \nu \rho
}^{\;\;\;\;\;\beta }-\frac{\sigma ^{\alpha \beta }}{2\cdot 4!}F^{\mu \nu
\rho \lambda }F_{\mu \nu \rho \lambda }  \label{stressen}
\end{equation}%
is the stress-energy tensor of the Abelian three-form gauge field. The
right-hand side of (\ref{4.19}) can be written like in the right-hand side
of (\ref{4.18}) if the following conditions are simultaneously satisfied
\begin{eqnarray}
-\delta \omega &=&-k\left( k+1\right) T_{\lambda }^{\pi }\left( \eta _{\xi
}\partial _{\left. {}\right. }^{[\lambda }h_{\pi }^{\xi ]}+\frac{1}{2}h_{\pi
\xi }\partial ^{\lbrack \lambda }\eta ^{\xi ]}\right) ,  \label{4.20} \\
\delta l^{\mu } &=&k\left( k+1\right) T_{\lambda }^{\mu }\eta _{\xi
}\partial ^{\lbrack \lambda }\eta ^{\xi ]}.  \label{4.21}
\end{eqnarray}%
Since none of the quantities $h_{\pi \xi }$, $\partial _{\left. {}\right.
}^{[\lambda }h_{\pi }^{\xi ]}$, $\eta _{\xi }$, or $\partial ^{\lbrack
\lambda }\eta ^{\xi ]}$ are $\delta $-exact, we deduce that the last
relations hold if stress-energy tensor of the Abelian three-form gauge field
is $\delta $-exact
\begin{equation}
T_{\tau }^{\mu }=\delta \Omega _{\tau }^{\mu }.  \label{4.22}
\end{equation}%
Assuming that the equation (\ref{4.22}) is valid, it further gives
\begin{equation}
\partial _{\mu }T_{\tau }^{\mu }=\delta \left( \partial _{\mu }\Omega _{\tau
}^{\mu }\right) .  \label{4.23}
\end{equation}%
On the other hand, by direct computation we find
\begin{equation}
\partial _{\mu }T_{\tau }^{\mu }=\delta \left( A^{\ast \nu \rho \lambda
}F_{\nu \rho \lambda \tau }\right) ,  \label{4.24}
\end{equation}%
so the right-hand side of (\ref{4.24}) cannot be written like in the
right-hand side of (\ref{4.23}). Therefore, relation (\ref{4.22}) is not
valid, and thus neither are (\ref{4.20})--(\ref{4.21}). As a consequence, $%
\chi $ must vanish, which further implies
\begin{equation}
k\left( k+1\right) =0.  \label{4.25}
\end{equation}%
The nontrivial solution to (\ref{4.25}) reads as (if we take $k=0$, then no
interactions occur)
\begin{equation}
k=-1.  \label{4.26}
\end{equation}%
Replacing (\ref{4.26}) in (\ref{wa2}) (and making use of (\ref{wa4})) and
then in (\ref{4.8})--(\ref{4.11})), we identify the components of the
second-order deformation as
\begin{equation}
b_{4}^{\mathrm{h-A}}=\frac{1}{2}C^{\ast }h_{\mu \nu }\eta ^{\mu }\partial
^{\nu }C,  \label{4.28}
\end{equation}%
\begin{eqnarray}
b_{3}^{\mathrm{h-A}} &=&\frac{1}{2}C^{\ast \mu }\left[ \frac{3}{4}h_{\mu
\rho }h_{\nu }^{\rho }\partial ^{\nu }C-\frac{1}{2}C^{\nu }\eta ^{\rho
}\partial _{\lbrack \mu }h_{\nu ]\rho }\right.  \notag \\
&&\left. +\frac{1}{4}C^{\nu }\left( h_{\mu }^{\rho }\partial _{\lbrack \nu
}\eta _{\rho ]}-h_{\nu }^{\rho }\partial _{\lbrack \mu }\eta _{\rho
]}\right) -h^{\nu \rho }\eta _{\rho }\partial _{\nu }C_{\mu }\right] ,
\label{4.30}
\end{eqnarray}%
\begin{eqnarray}
b_{2}^{\mathrm{h-A}} &=&-\frac{1}{2}C^{\ast \mu \nu }\left[ C^{\rho }h_{\nu
}^{\lambda }\partial _{\lbrack \mu }h_{\lambda ]\rho }+\frac{1}{2}C^{\rho
}\partial _{\mu }\left( h_{\nu \lambda }h_{\rho }^{\lambda }\right) -\frac{3%
}{2}h_{\mu \lambda }h^{\rho \lambda }\partial _{\rho }C_{\nu }\right.  \notag
\\
&&+\eta ^{\rho }\left( \partial _{\lbrack \mu }^{\left. {}\right. }h_{\rho
]}^{\lambda }-\partial _{\lbrack \xi }h_{\rho ]\mu }\sigma ^{\xi \lambda
}\right) C_{\nu \lambda }-\frac{1}{2}C_{\mu \rho }\left( h^{\rho \lambda
}\partial _{\lbrack \nu }\eta _{\lambda ]}-h_{\nu \lambda }\partial
^{\lbrack \rho }\eta ^{\lambda ]}\right)  \notag \\
&&\left. +\left( \partial _{\rho }C_{\mu \nu }\right) \eta _{\lambda
}h^{\rho \lambda }\right] ,  \label{4.32}
\end{eqnarray}%
\begin{eqnarray}
b_{1}^{\mathrm{h-A}} &=&\frac{3}{4}A^{\ast \mu \nu \rho }\left[ C_{\rho \xi
}\partial _{\mu }\left( h_{\nu \lambda }h^{\lambda \xi }\right) +\frac{3}{2}%
h_{\rho \xi }h^{\lambda \xi }\partial _{\lambda }C_{\mu \nu }+2C_{\mu
\lambda }h_{\nu }^{\xi }\partial _{\lbrack \xi }^{\left. {}\right. }h_{\rho
]}^{\lambda }\right.  \notag \\
&&-\frac{1}{2}A_{\mu \nu \lambda }\left( h^{\lambda \xi }\partial _{\lbrack
\rho }\eta _{\xi ]}+h_{\rho \xi }\partial ^{\lbrack \lambda }\eta ^{\xi
]}+2\sigma ^{\lambda \pi }\eta ^{\xi }\partial _{\lbrack \rho }h_{\pi ]\xi
}\right)  \notag \\
&&\left. +A_{\mu \nu \lambda }h_{\rho \xi }\partial ^{\lbrack \lambda }\eta
^{\xi ]}-\frac{2}{3}h^{\lambda \xi }\eta _{\xi }\partial _{\lambda }A_{\mu
\nu \rho }\right] ,  \label{4.34}
\end{eqnarray}%
and
\begin{eqnarray}
b_{0}^{\mathrm{h-A}} &=&\frac{1}{16}F^{\mu \nu \rho \lambda }F_{\mu \nu \xi
\pi }\left[ h_{\rho }^{\xi }h_{\lambda }^{\pi }-\frac{1}{3!}\delta _{\rho
}^{\xi }\delta _{\lambda }^{\pi }\left( \frac{1}{4}h^{2}-h^{\alpha \beta
}h_{\alpha \beta }\right) -\frac{1}{3}\delta _{\rho }^{\xi }h_{\lambda
\sigma }h^{\pi \sigma }\right]  \notag \\
&&+\frac{1}{16}F^{\mu \nu \rho \lambda }\left[ A_{\xi \rho \lambda }\partial
_{\mu }\left( h_{\nu \pi }h^{\lambda \pi }\right) -h_{\mu \pi }h^{\xi \pi
}\left( \partial _{\nu }A_{\xi \rho \lambda }+\frac{4}{3}\partial _{\xi
}A_{\nu \rho \lambda }\right) \right.  \notag \\
&&+A_{\mu \nu \xi }\left( 4h_{\rho }^{\pi }\partial _{\lbrack \pi }^{\left.
{}\right. }h_{\lambda ]}^{\xi }-h\partial _{\lbrack \rho }^{\left. {}\right.
}h_{\lambda ]}^{\xi }\right) -\frac{2}{3}h_{\lambda }^{\xi }h\partial _{\xi
}A_{\mu \nu \rho }-2A_{\mu \xi \pi }\partial _{\nu }\left( h_{\rho }^{\xi
}h_{\lambda }^{\pi }\right)  \notag \\
&&\left. +2h_{\rho }^{\xi }h_{\lambda }^{\pi }\partial _{\xi }A_{\pi \mu \nu
}\right] +\frac{1}{16}\partial _{\xi }\left( h_{[\mu }^{\pi }A_{\nu \rho
]\pi }^{\left. {}\right. }\right) \left[ \partial ^{\rho }\left( h_{\tau
}^{[\xi }A_{\left. {}\right. }^{\mu \nu ]\tau }\right) -\frac{1}{3}\partial
^{\xi }\left( h_{\tau }^{[\mu }A_{\left. {}\right. }^{\nu \rho ]\tau
}\right) \right]  \notag \\
&&+q\varepsilon ^{\mu _{1}\ldots \mu _{11}}\left( \frac{1}{2}hA_{\mu _{1}\mu
_{2}\mu _{3}}F_{\mu _{4}\ldots \mu _{7}}-4h_{\mu _{1}}^{\xi }A_{\mu _{2}\mu
_{3}\mu _{4}}F_{\mu _{5}\ldots \mu _{7}\xi }\right.  \notag \\
&&\left. +3h_{\mu _{1}}^{\xi }A_{\xi \mu _{2}\mu _{3}}F_{\mu _{4}\ldots \mu
_{7}}\right) F_{\mu _{8}\ldots \mu _{11}}.  \label{4.36}
\end{eqnarray}%
Formulas (\ref{4.28})--(\ref{4.36}) offer us the complete form of the
interacting part from the second-order deformation of the solution to the
master equation
\begin{equation}
S_{2}^{\mathrm{h-A}}=\int d^{11}x\left( b_{4}^{\mathrm{h-A}}+b_{3}^{\mathrm{%
h-A}}+b_{2}^{\mathrm{h-A}}+b_{1}^{\mathrm{h-A}}+b_{0}^{\mathrm{h-A}}\right) .
\label{4.37}
\end{equation}%
With the help of (\ref{4.26}), it results that $S_{1}^{\mathrm{h-A}}$ takes
the final form
\begin{eqnarray}
S_{1}^{\mathrm{h-A}} &=&\int d^{11}x\left\{ -\frac{1}{12}F^{\mu \nu \rho
\lambda }\left[ 3\partial _{\mu }\left( A_{\nu \rho \sigma }h_{\lambda
}^{\sigma }\right) -F_{\mu \nu \rho \sigma }h_{\lambda }^{\sigma }+\frac{1}{8%
}F_{\mu \nu \rho \lambda }h\right] \right.  \notag \\
&&+\frac{3}{2}A^{\ast \mu \nu \rho }\left( \frac{2}{3}\eta ^{\lambda
}\partial _{\lambda }A_{\mu \nu \rho }+A_{\mu \nu }^{\;\;\;\lambda }\partial
_{\lbrack \rho }\eta _{\lambda ]}-h_{\rho \lambda }\partial ^{\lambda
}C_{\mu \nu }-C_{\mu \lambda }\partial _{\lbrack \nu }^{\left. {}\right.
}h_{\rho ]}^{\lambda }\right)  \notag \\
&&+C^{\ast \mu \nu }\left[ \left( \partial _{\rho }C_{\mu \nu }\right) \eta
^{\rho }+C_{\mu }^{\;\;\rho }\partial _{\lbrack \nu }\eta _{\rho ]}+h_{\nu
\rho }\partial ^{\rho }C_{\mu }+\frac{1}{2}C^{\rho }\partial _{\lbrack \mu
}h_{\nu ]\rho }\right]  \notag \\
&&+\frac{1}{2}C^{\ast \mu }\left( 2\eta _{\nu }\partial ^{\nu }C_{\mu
}+C^{\nu }\partial _{\lbrack \mu }\eta _{\nu ]}-h_{\mu \nu }\partial ^{\nu
}C\right) +C^{\ast }\left( \partial ^{\mu }C\right) \eta _{\mu }  \notag \\
&&\left. +q\varepsilon ^{\mu _{1}\ldots \mu _{11}}A_{\mu _{1}\mu _{2}\mu
_{3}}F_{\mu _{4}\ldots \mu _{7}}F_{\mu _{8}\ldots \mu _{11}}\right\} .
\label{4.38}
\end{eqnarray}%
So far, we have completely determined the first- and second-order
deformations of the solution to the master equation corresponding to the
free model (\ref{fract}).

\section{Analysis of the deformed theory\label{defth}}

In Ref.~\cite{bizjhep} (Section 5) it has been shown that the local BRST
cohomologies of the Pauli-Fierz model and respectively of the linearized
version of vielbein formulation of spin-two field theory are isomorphic.
Because the local BRST cohomology (in ghost numbers zero and one) controls
the deformation procedure, it results that this isomorphism allows one to
pass in a consistent manner from the Pauli-Fierz version to the linearized
version of the vielbein formulation and conversely during the deformation
procedure. Nevertheless, the linearized vielbein formulation possesses more
fields (the antisymmetric part of the linearized vielbein) and more gauge
parameters (Lorentz parameters) than the Pauli-Fierz model, such that the
switch from the former version to the latter is realized via the above
mentioned isomorphism by imposing some partial gauge-fixing conditions,
which come from the more general ones~\cite{siegel}
\begin{equation}
\sigma _{\mu \lbrack a}^{\left. {}\right. }e_{b]}^{\;\;\mu }=0.  \label{gf}
\end{equation}%
In the context of the gauge-fixing conditions (\ref{gf}), simple computation
leads to the vielbein fields and their inverse up to the second order in the
coupling constant as
\begin{equation}
e_{a}^{\;\;\mu }=\overset{(0)}{e}_{a}^{\;\;\mu }+\lambda \overset{(1)}{e}%
_{a}^{\;\;\mu }+\lambda ^{2}\overset{(2)}{e}_{a}^{\;\;\mu }+\cdots =\delta
_{a}^{\;\;\mu }-\frac{\lambda }{2}h_{a}^{\;\;\mu }+\frac{3\lambda ^{2}}{8}%
h_{a}^{\;\;\rho }h_{\rho }^{\;\;\mu }+\cdots ,  \label{id1}
\end{equation}%
\begin{equation}
e_{\;\;\mu }^{a}=\overset{(0)}{e}_{\;\;\mu }^{a}+\lambda \overset{(1)}{e}%
_{\;\;\mu }^{a}+\lambda ^{2}\overset{(2)}{e}_{\;\;\mu }^{a}+\cdots =\delta
_{\;\;\mu }^{a}+\frac{\lambda }{2}h_{\;\;\mu }^{a}-\frac{\lambda ^{2}}{8}%
h_{\;\;\rho }^{a}h_{\;\;\mu }^{\rho }+\cdots .  \label{uv2a}
\end{equation}%
The first pieces from the expansion of the metric tensor and of its
determinant ($\sqrt{g}=\sqrt{\det g_{\mu \nu }}$) in terms of the
Pauli-Fierz field are written as
\begin{equation}
g^{\mu \nu }=\overset{(0)}{g^{\mu \nu }}+\lambda \overset{(1)}{g^{\mu \nu }}%
+\lambda ^{2}\overset{(2)}{g^{\mu \nu }}+\cdots =\sigma ^{\mu \nu }-\lambda
h^{\mu \nu }+\lambda ^{2}h_{\rho }^{\mu }h^{\rho \nu }+\cdots ,  \label{qqq1}
\end{equation}%
\begin{equation}
\sqrt{g}=\overset{(0)}{e=\sqrt{g}}+\lambda \overset{(1)}{\sqrt{g}}%
+\lambda ^{2}\overset{(2)}{\sqrt{g}}+\cdots =1+\frac{\lambda }{2}h+\frac{%
\lambda ^{2}}{8}\left( h^{2}-2h_{\mu \nu }h^{\mu \nu }\right) +\cdots ,
\label{a96}
\end{equation}%
where $e=\det e_{\;\;\mu }^{a}$.

Now, we have at hand all the ingredients required for the Lagrangian
formulation of the deformed theory obtained in the previous section. The
component of antighost number zero in $S_{1}^{\mathrm{h-A}}$ is precisely
the interacting Lagrangian at order one in the coupling constant
\begin{eqnarray}
\mathcal{L}_{1}^{\mathrm{h-A}} &=&-\frac{1}{12}F^{\mu \nu \rho \lambda
}\left( \frac{1}{8}F_{\mu \nu \rho \lambda }h-F_{\mu \nu \rho \sigma
}h_{\lambda }^{\sigma }+3\partial _{\mu }\left( A_{\nu \rho \sigma
}h_{\lambda }^{\sigma }\right) \right)  \notag \\
&&+q\varepsilon ^{\mu _{1}\ldots \mu _{11}}A_{\mu _{1}\mu _{2}\mu
_{3}}F_{\mu _{4}\ldots \mu _{7}}F_{\mu _{8}\ldots \mu _{11}}.  \label{a93}
\end{eqnarray}%
It can be put under the more suggestive form
\begin{eqnarray}
\mathcal{L}_{1}^{\mathrm{h-A}} &=&-\frac{1}{2\cdot 4!}\overset{\left(
0\right) }{g}^{\mu \alpha }\overset{\left( 0\right) }{g}^{\nu \beta }\overset%
{\left( 0\right) }{g}^{\rho \gamma }\overset{\left( 0\right) }{\bar{F}}_{\mu
\nu \rho \lambda }\left( \overset{(1)}{\sqrt{g}}\overset{\left( 0\right) }{g%
}^{\lambda \delta }\overset{\left( 0\right) }{\bar{F}}_{\alpha \beta \gamma
\delta }\right.  \notag \\
&&\left. +4\overset{(0)}{\sqrt{g}}\overset{\left( 1\right)
}{g}^{\lambda
\delta }\overset{\left( 0\right) }{\bar{F}}_{\alpha \beta \gamma \delta }+2%
\overset{(0)}{\sqrt{g}}\overset{\left( 0\right) }{g}^{\lambda \delta }%
\overset{\left( 1\right) }{\bar{F}}_{\alpha \beta \gamma \delta }\right)
\notag \\
&&+q\overset{(0)}{\sqrt{g}}\overset{\left( 0\right)
}{e}_{a_{1}}^{\;\;\mu _{1}}\cdots \overset{\left( 0\right)
}{e}_{a_{11}}^{\;\;\mu _{11}}\epsilon ^{a_{1}\ldots
a_{11}}\overset{\left( 0\right) }{\bar{A}}_{\mu _{1}\mu
_{2}\mu _{3}}\overset{\left( 0\right) }{\bar{F}}_{\mu _{4}\ldots \mu _{7}}%
\overset{\left( 0\right) }{\bar{F}}_{\mu _{8}\ldots \mu _{11}},  \label{a93a}
\end{eqnarray}%
where
\begin{eqnarray}
\overset{\left( 0\right) }{\bar{A}}_{\mu \nu \rho } &=&\overset{\left(
0\right) }{e}_{\;\;\mu }^{a}\overset{\left( 0\right) }{e}_{\;\;\nu }^{b}%
\overset{\left( 0\right) }{e}_{\;\;\rho }^{c}A_{abc},\qquad \overset{\left(
0\right) }{\bar{F}}_{\mu \nu \rho \lambda }\equiv \partial _{\lbrack \mu }%
\overset{\left( 0\right) }{\bar{A}}_{\nu \rho \lambda ]},  \label{a96a} \\
\overset{\left( 1\right) }{\bar{A}}_{\mu \nu \rho } &=&\overset{\left(
1\right) }{e}_{\;\;[\mu }^{a}\overset{\left( 0\right) }{e}_{\;\;\nu }^{b}%
\overset{\left( 0\right) }{e}_{\;\;\rho] }^{c}A_{abc},\qquad \overset{\left(
1\right) }{\bar{F}}_{\mu \nu \rho \lambda }\equiv \partial _{\lbrack \mu }%
\overset{\left( 1\right) }{\bar{A}}_{\nu \rho \lambda ]}.  \label{a96w}
\end{eqnarray}%
In the first formula from equation (\ref{a96a}) $A_{abc}$ is nothing but the
original three-form gauge field (with flat indices). Along the same line,
the piece of antighost number equal to zero from the second-order
deformation furnishes us (up to a total derivative) with the interacting
Lagrangian at order two in the coupling constant
\begin{eqnarray}
\mathcal{L}_{2}^{\mathrm{h-A}} &=&-\frac{1}{2\cdot 4!}\overset{\left(
0\right) }{g}^{\mu \alpha }\overset{\left( 0\right) }{g}^{\nu \beta }\left[
\overset{\left( 0\right) }{\bar{F}}_{\mu \nu \rho \lambda }\left( \overset{%
(2)}{\sqrt{g}}\overset{\left( 0\right) }{g}^{\rho \gamma
}\overset{\left( 0\right) }{g}^{\lambda \delta }\overset{\left(
0\right) }{\bar{F}}_{\alpha
\beta \gamma \delta }\right. \right.  \notag \\
&&+4\overset{(0)}{\sqrt{g}}\overset{\left( 2\right) }{g}^{\rho \gamma }%
\overset{\left( 0\right) }{g}^{\lambda \delta }\overset{\left( 0\right) }{%
\bar{F}}_{\alpha \beta \gamma \delta }+2\overset{(0)}{\sqrt{g}}\overset{%
\left( 0\right) }{g}^{\rho \gamma }\overset{\left( 0\right) }{g}^{\lambda
\delta }\overset{\left( 2\right) }{\bar{F}}_{\alpha \beta \gamma \delta }
\notag \\
&&+4\overset{(1)}{\sqrt{g}}\overset{\left( 1\right) }{g}^{\rho \gamma }%
\overset{\left( 0\right) }{g}^{\lambda \delta }\overset{\left( 0\right) }{%
\bar{F}}_{\alpha \beta \gamma \delta }+2\overset{(1)}{\sqrt{g}}\overset{%
\left( 0\right) }{g}^{\rho \gamma }\overset{\left( 0\right) }{g}^{\lambda
\delta }\overset{\left( 1\right) }{\bar{F}}_{\alpha \beta \gamma \delta }
\notag \\
&&\left. +6\overset{(0)}{\sqrt{g}}\overset{\left( 1\right)
}{g}^{\rho \gamma }\overset{\left( 1\right) }{g}^{\lambda \delta
}\overset{\left(
0\right) }{\bar{F}}_{\alpha \beta \gamma \delta }+8\overset{(0)}{\sqrt{g}}%
\overset{\left( 1\right) }{g}^{\rho \gamma }\overset{\left( 0\right) }{g}%
^{\lambda \delta }\overset{\left( 1\right) }{\bar{F}}_{\alpha \beta \gamma
\delta }\right)  \notag \\
&&\left. +\overset{(0)}{\sqrt{g}}\overset{\left( 0\right) }{g}^{\rho
\gamma
}\overset{\left( 0\right) }{g}^{\lambda \delta }\overset{\left( 1\right) }{%
\bar{F}}_{\mu \nu \rho \lambda }\overset{\left( 1\right) }{\bar{F}}_{\alpha
\beta \gamma \delta }\right] +q\epsilon ^{a_{1}\ldots a_{11}}\overset{\left(
0\right) }{e}_{a_{1}}^{\;\;\mu _{1}}\cdots \overset{\left( 0\right) }{e}%
_{a_{10}}^{\;\;\mu _{10}}\overset{\left( 0\right) }{\bar{F}}_{\mu _{4}\ldots
\mu _{7}}  \notag \\
&&\times \left[ \left( \overset{\left( 1\right)
}{\sqrt{g}}\overset{\left(
0\right) }{e}_{a_{11}}^{\;\;\mu _{11}}+8\overset{\left( 0\right) }{\sqrt{g}}%
\overset{\left( 1\right) }{e}_{a_{11}}^{\;\;\mu _{11}}\right) \overset{%
\left( 0\right) }{\bar{A}}_{\mu _{1}\mu _{2}\mu _{3}}\overset{\left(
0\right) }{\bar{F}}_{\mu _{8}\ldots \mu _{11}}\right.  \notag \\
&&\left. +\overset{\left( 0\right) }{\sqrt{g}}\overset{\left( 0\right) }{e}%
_{a_{11}}^{\;\;\mu _{11}}\left( \overset{\left( 1\right) }{\bar{A}}_{\mu
_{1}\mu _{2}\mu _{3}}\overset{\left( 0\right) }{\bar{F}}_{\mu _{8}\ldots \mu
_{11}}+2\overset{\left( 0\right) }{\bar{A}}_{\mu _{1}\mu _{2}\mu _{3}}%
\overset{\left( 1\right) }{\bar{F}}_{\mu _{8}\ldots \mu _{11}}\right) \right]
\notag \\
&&+3q\epsilon ^{a_{1}\ldots a_{11}}\overset{\left( 1\right) }{e}%
_{a_{1}}^{\;\;\mu _{1}}\overset{\left( 0\right) }{e}_{a_{2}}^{\;\;\mu
_{2}}\cdots \overset{\left( 0\right) }{e}_{a_{11}}^{\;\;\mu _{11}}\overset{%
\left( 0\right) }{\bar{A}}_{\mu _{1}\mu _{2}\mu _{3}}\overset{\left(
0\right) }{\bar{F}}_{\mu _{4}\ldots \mu _{7}}\overset{\left( 0\right) }{\bar{%
F}}_{\mu _{8}\ldots \mu _{11}}.  \label{a98}
\end{eqnarray}%
With the help of (\ref{a93}) and (\ref{a98}) we deduce that $\mathcal{L}%
_{0}^{\mathrm{A}}+\lambda \mathcal{L}_{1}^{\mathrm{h-A}}+\lambda ^{2}%
\mathcal{L}_{2}^{\mathrm{h-A}}+\cdots $ comes from the expansion of the
fully deformed Lagrangian
\begin{equation}
\mathcal{L}^{\mathrm{h-A}}=-\frac{1}{2\cdot 4!}\sqrt{g}\bar{F}_{\mu
\nu \rho \lambda }\bar{F}^{\mu \nu \rho \lambda }+\lambda q\epsilon
^{\mu _{1}\ldots \mu _{11}}\bar{A}_{\mu _{1}\mu _{2}\mu
_{3}}\bar{F}_{\mu _{4}\ldots \mu _{7}}\bar{F}_{\mu _{8}\ldots \mu
_{11}},  \label{a99}
\end{equation}%
where
\begin{eqnarray}
\bar{F}_{\mu \nu \rho \lambda } &=&\partial _{\lbrack \mu }\left( e_{\;\;\nu
}^{a}e_{\;\;\rho }^{b}e_{\;\;\lambda ]}^{c}A_{abc}\right) ,
\label{notFbarjos} \\
\bar{F}^{\mu \nu \rho \lambda } &=&g^{\mu \alpha }g^{\nu \beta }g^{\rho
\gamma }g^{\lambda \delta }\bar{F}_{\alpha \beta \gamma \delta },
\label{notFbarsus} \\
\epsilon ^{\mu _{1}\ldots \mu _{11}} &=&\sqrt{g}e_{a_{1}}^{\;\;\mu
_{1}}e_{a_{2}}^{\;\;\mu _{2}}\cdots e_{a_{11}}^{\;\;\mu
_{11}}\epsilon ^{a_{1}\ldots a_{11}}.  \label{noteps}
\end{eqnarray}

The pieces from the deformed solution to the master equation that are linear
in the antifields $A^{\ast \alpha \beta \gamma }$\ produce the deformed
gauge transformations of the Abelian three-form gauge field
\begin{eqnarray}
\bar{\delta}_{\epsilon ,\varepsilon }A_{\alpha \beta \gamma } &=&\partial
_{\lbrack \alpha }\varepsilon _{\beta \gamma ]}+\lambda \left[ \epsilon
^{\delta }\partial _{\delta }A_{\alpha \beta \gamma }+\frac{1}{2}%
A_{\;\;[\alpha \beta }^{\delta }\delta _{\gamma ]}^{\sigma }\partial
_{\lbrack \sigma }\epsilon _{\delta ]}\right.  \notag \\
&&\left. -\frac{1}{2}\left( \partial ^{\delta }\varepsilon _{\lbrack \alpha
\beta }\right) h_{\gamma ]\delta }+\frac{1}{2}\varepsilon _{\;\;[\alpha
}^{\delta }\partial _{\beta }^{\left. {}\right. }h_{\gamma ]\delta }^{\left.
{}\right. }\right]  \notag \\
&&+\lambda ^{2}\left[ -\frac{1}{8}\varepsilon _{\;\;[\alpha }^{\delta
}\left( \partial _{\beta }^{\left. {}\right. }h_{\gamma ]}^{\sigma }\right)
h_{\delta \sigma }+\frac{3}{8}\varepsilon _{\;\;[\alpha }^{\delta }h_{\beta
}^{\sigma }\partial _{\gamma ]}^{\left. {}\right. }h_{\delta \sigma }\right.
\notag \\
&&+\frac{3}{8}\left( \partial ^{\delta }\varepsilon _{\lbrack \alpha \beta
}\right) h_{\gamma ]}^{\sigma }h_{\delta \sigma }-\frac{1}{4}\left( \partial
_{\delta }h_{[\alpha }^{\sigma }\right) h_{\beta }^{\delta }\varepsilon
_{\gamma ]\sigma }  \notag \\
&&-\frac{1}{8}A_{\;\;[\alpha \beta }^{\delta }\delta _{\gamma ]}^{\omega
}\left( \partial _{\lbrack \omega }\epsilon _{\sigma ]}\right) h_{\delta
}^{\sigma }+\frac{1}{8}A_{\;\;[\alpha \beta }^{\delta }h_{\gamma ]}^{\sigma
}\partial _{\lbrack \delta }\epsilon _{\sigma ]}  \notag \\
&&\left. -\frac{1}{4}A_{\;\;[\alpha \beta }^{\delta }\delta _{\gamma
]}^{\omega }\partial \left( _{\lbrack \omega }h_{\delta ]}^{\sigma }\right)
\epsilon _{\sigma }-\frac{1}{2}\left( \partial ^{\delta }A_{\alpha \beta
\gamma }\right) h_{\delta }^{\sigma }\epsilon _{\sigma }\right] +\cdots
\notag \\
&=&\overset{\left( 0\right) }{\overline{\delta }}_{\epsilon ,\varepsilon
}A_{\alpha \beta \gamma }+\lambda \overset{\left( 1\right) }{\overline{%
\delta }}_{\epsilon ,\varepsilon }A_{\alpha \beta \gamma }+\lambda ^{2}%
\overset{\left( 2\right) }{\overline{\delta }}_{\epsilon ,\varepsilon
}A_{\alpha \beta \gamma }+\cdots .  \label{a100}
\end{eqnarray}%
We recall that the initial three-form gauge field possesses flat indices,
i.e. $A_{\alpha \beta \gamma }$ means $A_{abc}$. The contributions of orders
one and two to the above gauge transformations can be put under the form
\begin{equation}
\overset{\left( 1\right) }{\bar{\delta}}_{\epsilon ,\varepsilon }A_{abc}=%
\overset{\left( 0\right) }{\bar{\epsilon}}^{\mu }\partial _{\mu
}A_{abc}+A_{\;\;[ab}^{m}\overset{\left( 0\right) }{\epsilon }_{c]m}^{\left.
{}\right. }+\left( \partial _{\mu }\varepsilon _{\lbrack ab}\right) \overset{%
\left( 1\right) }{e}_{c]}^{\;\;\mu }+\frac{1}{2}\overset{\left( 0\right) }{e}%
_{m}^{\;\;\mu }\overset{\left( 1\right) }{\omega }_{\mu \lbrack ab}^{\left.
{}\right. }\varepsilon _{c]}^{\;\;m},  \label{def2}
\end{equation}%
\begin{eqnarray}
\overset{\left( 2\right) }{\bar{\delta}}_{\epsilon ,\varepsilon }A_{abc} &=&%
\overset{\left( 1\right) }{\bar{\epsilon}}^{\mu }\partial _{\mu
}A_{abc}+A_{\;\;[ab}^{m}\overset{\left( 1\right) }{\epsilon }_{c]m}^{\left.
{}\right. }+\left( \partial _{\mu }\varepsilon _{\lbrack ab}\right) \overset{%
\left( 2\right) }{e}_{c]}^{\;\;\mu }  \notag \\
&&+\frac{1}{2}\overset{\left( 1\right) }{e}_{m}^{\;\;\mu }\overset{\left(
1\right) }{\omega }_{\mu \lbrack ab}^{\left. {}\right. }\varepsilon
_{c]}^{\;\;m}+\frac{1}{2}\overset{\left( 0\right) }{e}_{m}^{\;\;\mu }\overset%
{\left( 2\right) }{\omega }_{\mu \lbrack ab}^{\left. {}\right. }\varepsilon
_{c]}^{\;\;m},  \label{def3}
\end{eqnarray}%
where we used the notations
\begin{eqnarray}
\overset{(0)}{\bar{\epsilon}}^{\mu } &=&\epsilon ^{\mu }=\epsilon ^{a}\delta
_{a}^{\;\;\mu },\qquad \overset{(1)}{\bar{\epsilon}}^{\mu }=-\frac{1}{2}%
\epsilon ^{a}h_{a}^{\;\;\mu },  \label{uv16} \\
\overset{(0)}{\epsilon }_{ab} &=&\frac{1}{2}\partial _{\lbrack a}\epsilon
_{b]},  \label{apx0} \\
\overset{(1)}{\epsilon }_{ab} &=&-\frac{1}{4}\epsilon ^{c}\partial _{\lbrack
a}h_{b]c}+\frac{1}{8}h_{[a}^{c}\partial _{b]}^{\left. {}\right. }\epsilon
_{c}+\frac{1}{8}\left( \partial _{c}\epsilon _{\lbrack a}^{\left. {}\right.
}\right) h_{b]}^{c},  \label{apx1} \\
\overset{\left( 1\right) }{\omega }_{\mu ab} &=&-\partial _{\lbrack
a}h_{b]\mu },  \label{c1} \\
\overset{\left( 2\right) }{\omega }_{\mu ab} &=&-\frac{1}{4}\left(
2h_{c[a}\left( \partial _{b]}h_{\;\;\mu }^{c}\right) -2h_{\left[ a\right.
}^{\;\;\;\nu }\partial _{\nu }h_{\left. b\right] \mu }-\left( \partial _{\mu
}h_{[a}^{\;\;\;\nu }\right) h_{b]\nu }\right) .  \label{c2}
\end{eqnarray}%
In formulas (\ref{apx0}) and (\ref{apx1}) the gauge parameters $\overset{(0)}%
{\epsilon }_{ab}$ and $\overset{(1)}{\epsilon }_{ab}$ are precisely the
first two terms from the Lorentz parameters expressed in terms of the flat
parameters $\epsilon ^{a}$ via the partial gauge fixing (\ref{gf}). Indeed, (%
\ref{gf}) leads to
\begin{equation}
\delta _{\epsilon }\left( \sigma _{\mu \lbrack a}^{\left. {}\right.
}e_{b]}^{\;\;\mu }\right) =0,  \label{vgf}
\end{equation}%
where
\begin{equation}
\frac{1}{\lambda }\delta _{\epsilon }e_{a}^{\;\;\mu }=\bar{\epsilon}^{\rho
}\partial _{\rho }e_{a}^{\;\;\mu }-e_{a}^{\;\;\rho }\partial _{\rho }\bar{%
\epsilon}^{\mu }+\epsilon _{a}^{\;\;b}e_{b}^{\;\;\mu }.  \label{id6}
\end{equation}%
Substituting (\ref{id1}) together with the expansions
\begin{equation}
\bar{\epsilon}^{\mu }=\overset{(0)}{\bar{\epsilon}}^{\mu }+\lambda \overset{%
(1)}{\bar{\epsilon}}^{\mu }+\cdots =\left( \delta _{a}^{\;\;\mu }-\frac{%
\lambda }{2}h_{a}^{\;\;\mu }+\cdots \right) \epsilon ^{a}  \label{uv15}
\end{equation}%
and
\begin{equation}
\epsilon _{ab}=\overset{(0)}{\epsilon }_{ab}+\lambda \overset{(1)}{\epsilon }%
_{ab}+\cdots  \label{uv12}
\end{equation}%
in (\ref{vgf}), we arrive precisely to (\ref{apx0})--(\ref{apx1}). In
formulas (\ref{c1}) and (\ref{c2}) $\overset{\left( 1\right) }{\omega }_{\mu
ab}$ and $\overset{\left( 2\right) }{\omega }_{\mu ab}$ represent the first-
and respectively second-order approximation of the spin connection
\begin{eqnarray}
\omega _{\mu ab} &=&e_{b}^{\;\;\nu }\partial _{\nu }e_{a\mu }-e_{a}^{\;\;\nu
}\partial _{\nu }e_{b\mu }+e_{a\nu }\partial _{\mu }e_{b}^{\;\;\nu }  \notag
\\
&&-e_{b\nu }\partial _{\mu }e_{a}^{\;\;\nu }+e_{\left[ a\right. }^{\;\;\rho
}e_{\left. b\right] }^{\;\;\nu }e_{c\mu }\partial _{\nu }e_{\;\;\rho }^{c}
\notag \\
&=&\lambda \overset{\left( 1\right) }{\omega }_{\mu ab}+\lambda ^{2}\overset{%
\left( 2\right) }{\omega }_{\mu ab}+\cdots .  \label{uv2}
\end{eqnarray}%
At this point it is easy to see that the deformed gauge transformations of
the three-form gauge field (see formula (\ref{a100})) come from the
perturbative expansion of the full gauge transformations
\begin{equation}
\bar{\delta}_{\epsilon ,\varepsilon }A_{abc}=\lambda \left( \bar{\epsilon}%
^{\mu }\partial _{\mu }A_{abc}+A_{\;\;[ab}^{m}\epsilon _{c]m}^{\left.
{}\right. }\right) +\left( \partial _{\mu }\varepsilon _{\lbrack ab}^{\left.
{}\right. }\right) e_{c]}^{\;\;\mu }+\frac{1}{2}e_{m}^{\;\;\mu }\omega _{\mu
\lbrack ab}^{\left. {}\right. }\varepsilon _{c]}^{\;\;m}.  \label{trflt}
\end{equation}%
The gauge transformations of the three-form with curved indices are obtained
with the help of (\ref{id6}) and (\ref{trflt})
\begin{equation}
\bar{\delta}_{\bar{\varepsilon},\bar{\epsilon}}\bar{A}_{\mu \nu \rho
}=\partial _{\lbrack \mu }\bar{\varepsilon}_{\nu \rho ]}+\lambda \left( \bar{%
\epsilon}^{\lambda }\partial _{\lambda }\bar{A}_{\mu \nu \rho }+\bar{A}%
_{\sigma \lbrack \mu \nu }\partial _{\rho ]}\bar{\epsilon}^{\sigma }\right) ,
\label{tf2}
\end{equation}%
where
\begin{equation}
\bar{\varepsilon}_{\mu \nu }=e_{\;\;\mu }^{a}e_{\;\;\nu }^{b}\varepsilon
_{ab}.  \label{parc}
\end{equation}

We observe that (\ref{tf2}) describes a set of gauge transformations that
remain off-shell, second-order reducible. Indeed, if we make the
transformations
\begin{equation}
\bar{\varepsilon}_{\mu \nu }\rightarrow \bar{\varepsilon}_{\mu \nu }^{\left(
\bar{\theta}\right) }=\partial _{\lbrack \mu }\bar{\theta}_{\nu ]},
\label{f3def}
\end{equation}%
then the gauge variation of the three-form identically vanishes
\begin{equation}
\bar{\delta}_{\bar{\varepsilon}^{\left( \bar{\theta}\right) }}\bar{A}_{\mu
\nu \rho }\equiv 0.  \label{f3adef}
\end{equation}%
Moreover, if in (\ref{f3def}) we perform the changes
\begin{equation}
\bar{\theta}_{\mu }\rightarrow \bar{\theta}_{\mu }^{\left( \phi \right)
}=\partial _{\mu }\phi ,  \label{f4def}
\end{equation}%
with $\phi $\ an arbitrary scalar field, then the transformed gauge
parameters (\ref{f3def}) identically vanish
\begin{equation}
\bar{\varepsilon}_{\mu \nu }^{\left( \bar{\theta}^{\left( \phi \right)
}\right) }\equiv 0.  \label{f4adef}
\end{equation}%
The results concerning the reducibility relations for the interacting theory
can be read from the pieces that are simultaneously linear in the ghosts and
in the antifields (with the antighost number equal to two or three from the
deformed solution to the master equation).

In conclusion, under the hypotheses mentioned at the beginning of subsection %
\ref{stand}, we obtained that a candidate to the Lagrangian responsible for
the interactions between the spin-two field and a three-form gauge field in $%
D=11$ is described in (\ref{a99}) and the deformed gauge transformations of
the three-form are given by (\ref{tf2}).

\section{Uniqueness of interactions\label{unique}}

So far, we emphasized that there exists one candidate describing the
consistent interactions between one graviton and an Abelian three-form gauge
field, namely
\begin{eqnarray}
\tilde{\mathcal{L}}&=&\frac{2}{\lambda ^{2}}e\left( R-2\lambda ^{2}\Lambda
\right) -\frac{1}{2\cdot 4!}e\bar{F}_{\mu \nu \rho \lambda }\bar{F}^{\mu \nu
\rho \lambda }  \notag \\
&&+\lambda q\varepsilon ^{\mu _{1}\mu _{2}\cdots \mu _{11}}\bar{A}_{\mu
_{1}\mu _{2}\mu _{3}}\bar{F}_{\mu _{4}\cdots \mu _{7}}\bar{F}_{\mu
_{8}\cdots \mu _{11}},  \label{1}
\end{eqnarray}
in the context of the partial gauge-fixing (\ref{vgf}). So, the only point
that remains to be done is to check that there are no other solutions.

Let us denote by $\tilde{S}$ the solution to the master equation for the
theory with the standard Lagrangian (\ref{1}) decomposed according to the
power orders of the coupling constant $\lambda $
\begin{equation}
\tilde{S}=\tilde{S}_{0}+\lambda \tilde{S}_{1}+\lambda ^{2}\tilde{S}%
_{2}+\lambda ^{3}\tilde{S}_{3}+\lambda ^{4}\tilde{S}_{4}+\cdots  \label{ein}
\end{equation}
and by $S$ the fully deformed solution of the master equation associated
with the free theory (\ref{fract}), consistent to all orders in the coupling
constant
\begin{equation}
S=\bar{S}+\lambda S_{1}+\lambda ^{2}S_{2}+\lambda ^{3}S_{3}+\cdots ,
\label{our}
\end{equation}
such that they respectively fulfill the equations
\begin{eqnarray}
\left( \tilde{S},\tilde{S}\right) &=&0,  \label{maststilde} \\
\left( S,S\right) &=&0.  \label{masts}
\end{eqnarray}
Until now we investigated $\bar{S}$, $S_{1}$, and $S_{2}$ and proved that
they coincide with the standard ones
\begin{equation}
\bar{S}=\tilde{S}_{0},\qquad S_{1}=\tilde{S}_{1},\qquad S_{2}=\tilde{S}_{2}
\label{un0}
\end{equation}
in the presence of the partial gauge-fixing (\ref{vgf}). The question is how
unique are $S_{3}$, $S_{4}$, etc. given (\ref{our}). We will answer this
question by showing that the interactions provided by our deformation
procedure can always be brought to those prescribed by the usual rules from
General Relativity via a suitable redefinition of the constants $\lambda $, $%
q$, and $\Lambda $ from (\ref{1}). More precisely, we will prove that the
fully deformed solution (\ref{our}) is nothing but (\ref{ein}) up to the
replacements
\begin{eqnarray}
\lambda &\rightarrow &\lambda \left( 1+k_{3}^{(1)}\lambda
^{2}+k_{4}^{(1)}\lambda ^{3}+k_{5}^{(1)}\lambda ^{4}+\cdots \right) ,
\label{deflambda} \\
\Lambda &\rightarrow &\Lambda \frac{1+k_{3}^{(4)}\lambda
^{2}+k_{4}^{(4)}\lambda ^{3}+k_{5}^{(4)}\lambda ^{4}+\cdots }{%
1+k_{3}^{(1)}\lambda ^{2}+k_{4}^{(1)}\lambda ^{3}+k_{5}^{(1)}\lambda
^{4}+\cdots },  \label{defLambda} \\
q &\rightarrow &q\frac{1+k_{3}^{(3)}\lambda ^{2}+k_{4}^{(3)}\lambda
^{3}+k_{5}^{(3)}\lambda ^{4}+\cdots }{1+k_{3}^{(1)}\lambda
^{2}+k_{4}^{(1)}\lambda ^{3}+k_{5}^{(1)}\lambda ^{4}+\cdots },  \label{defq}
\end{eqnarray}
with $k_{j}^{(m)}$ some arbitrary, real constants.

Our starting point is that (\ref{ein}) and (\ref{our}) respectively satisfy
equations (\ref{maststilde}) and (\ref{masts}) together with relations (\ref%
{un0}). The projection of (\ref{ein}) and (\ref{our}) on $\lambda ^{3}$
emphasizes that $S_{3}$ and respectively $\tilde{S}_{3}$ are solutions to
the equations
\begin{equation}
sS_{3}=-\left( S_{1},S_{2}\right) ,\qquad s\tilde{S}_{3}=-\left( \tilde{S}%
_{1},\tilde{S}_{2}\right) .  \label{un1}
\end{equation}%
Recalling (\ref{un0}) and subtracting the latter equation in (\ref{un1})
from the former we obtain
\begin{equation}
s\left( S_{3}-\tilde{S}_{3}\right) =0,  \label{eqs3}
\end{equation}%
whose general solution, according to our results from subsection \ref%
{firstord} (and to the second equality from (\ref{un0})), reads as
\begin{equation}
S_{3}-\tilde{S}_{3}=\sum\limits_{m=1}^{4}k_{3}^{(m)}\tilde{S}_{1}^{(m)}.
\label{un3}
\end{equation}%
In the above $\left( k_{3}^{(m)}\right) _{m=\overline{1,4}}$ are arbitrary,
real constants and $\left( \tilde{S}_{1}^{(m)}\right) _{m=\overline{1,4}}$
are the independent components of the first-order deformation $S_{1}=\tilde{S%
}_{1}$ (they individually satisfy the equation $s\tilde{S}_{1}^{(m)}=0$)
\begin{equation}
\tilde{S}_{1}=\tilde{S}_{1}^{(1)}+\tilde{S}_{1}^{(2)}+\tilde{S}_{1}^{(3)}+%
\tilde{S}_{1}^{(4)},  \label{dezv1}
\end{equation}%
namely, $\tilde{S}_{1}^{(1)}$ represents the first-order deformation of the
solution to the master equation from the Pauli-Fierz sector containing the
cubic vertex of the Einstein-Hilbert Lagrangian (see (\ref{3.12w})), but not
the cosmological term, $\tilde{S}_{1}^{(2)}$ denotes the interacting part of
the first-order deformation (see (\ref{fo39}) for $k=-1$), $\tilde{S}%
_{1}^{(3)}$ stands for the first-order deformation in the three-form sector
(see (\ref{fo5})), linear in $q$, and $\tilde{S}_{1}^{(4)}$ means the
first-order deformation from the Pauli-Fierz sector that does not modify the
gauge transformations of the graviton (the cosmological term, linear in the
cosmological constant $\Lambda $). By direct computation we find that the
various antibrackets among $\tilde{S}_{1}^{(m)}$ read as
\begin{gather}
\left( \tilde{S}_{1}^{(1)},\tilde{S}_{1}^{(1)}\right) =-2s\tilde{S}%
_{2}^{(1)},\quad \left( \tilde{S}_{1}^{(1)},\tilde{S}_{1}^{(3)}\right)
=0,\quad \left( \tilde{S}_{1}^{(1)},\tilde{S}_{1}^{(4)}\right) =-s\tilde{S}%
_{2}^{(4)},  \label{p1} \\
\left( \tilde{S}_{1}^{(2)},\tilde{S}_{1}^{(2)}\right) +2\left( \tilde{S}%
_{1}^{(1)},\tilde{S}_{1}^{(2)}\right) =-2s\tilde{S}_{2}^{(2)},\qquad \left(
\tilde{S}_{1}^{(2)},\tilde{S}_{1}^{(3)}\right) =-s\tilde{S}_{2}^{(3)},
\label{p2} \\
\left( \tilde{S}_{1}^{(2)},\tilde{S}_{1}^{(4)}\right) =\left( \tilde{S}%
_{1}^{(3)},\tilde{S}_{1}^{(3)}\right) =\left( \tilde{S}_{1}^{(3)},\tilde{S}%
_{1}^{(4)}\right) =\left( \tilde{S}_{1}^{(4)},\tilde{S}_{1}^{(4)}\right) =0,
\label{p3}
\end{gather}%
where $\left( \tilde{S}_{2}^{(m)}\right) _{m=\overline{1,4}}$ are the
components of the second-order deformation of the solution to the master
equation $S_{2}=\tilde{S}_{2}$ (see (\ref{4.2}) and (\ref{4.37})), $\left(
\tilde{S}_{1},\tilde{S}_{1}\right) =-2s\tilde{S}_{2}$, respectively induced
by the decomposition (\ref{dezv1})
\begin{equation}
\tilde{S}_{2}=\tilde{S}_{2}^{(1)}+\tilde{S}_{2}^{(2)}+\tilde{S}_{2}^{(3)}+%
\tilde{S}_{2}^{(4)}.  \label{dezv2}
\end{equation}%
Based on the concrete form of the various components from (\ref{dezv1}) and (%
\ref{dezv2}), it can be shown that their antibrackets can be expressed as
\begin{gather}
\left( \tilde{S}_{1}^{(1)},\tilde{S}_{2}^{(1)}\right) =-s\tilde{S}%
_{3}^{(1)},\;\left( \tilde{S}_{1}^{(1)},\tilde{S}_{2}^{(4)}\right) +\left(
\tilde{S}_{1}^{(4)},\tilde{S}_{2}^{(1)}\right) =-s\tilde{S}_{3}^{(4)},
\label{q1} \\
\left( \tilde{S}_{1}^{(1)},\tilde{S}_{2}^{(2)}\right) +\left( \tilde{S}%
_{2}^{(1)},\tilde{S}_{1}^{(2)}\right) +\left( \tilde{S}_{1}^{(2)},\tilde{S}%
_{2}^{(2)}\right) =-s\tilde{S}_{3}^{(2)},  \label{q2} \\
\left( \tilde{S}_{1}^{(1)},\tilde{S}_{2}^{(3)}\right) +\left( \tilde{S}%
_{1}^{(3)},\tilde{S}_{2}^{(1)}\right) +\left( \tilde{S}_{1}^{(2)},\tilde{S}%
_{2}^{(3)}\right) +\left( \tilde{S}_{1}^{(3)},\tilde{S}_{2}^{(2)}\right) =-s%
\tilde{S}_{3}^{(3)},  \label{q3} \\
\left( \tilde{S}_{1}^{(3)},\tilde{S}_{2}^{(3)}\right) =\left( \tilde{S}%
_{1}^{(3)},\tilde{S}_{2}^{(4)}\right) =\left( \tilde{S}_{1}^{(4)},\tilde{S}%
_{2}^{(3)}\right) =\left( \tilde{S}_{1}^{(4)},\tilde{S}_{2}^{(4)}\right) =0,
\label{q4}
\end{gather}%
where $\left( \tilde{S}_{3}^{(m)}\right) _{m=\overline{1,4}}$ are the
components of the solution to the master equation for the theory with the
standard Lagrangian (\ref{1}) of order three in the coupling constant, $%
\tilde{S}_{3}$,
\begin{equation}
\tilde{S}_{3}=\tilde{S}_{3}^{(1)}+\tilde{S}_{3}^{(2)}+\tilde{S}_{3}^{(3)}+%
\tilde{S}_{3}^{(4)}.  \label{dezv3}
\end{equation}

At the same time, the various terms from (\ref{dezv1}), (\ref{dezv2}), and (%
\ref{dezv3}) check the individual equations
\begin{eqnarray}
&&\left( \tilde{S}_{2}^{(1)},\tilde{S}_{2}^{(1)}\right) +2\left( \tilde{S}%
_{1}^{(1)},\tilde{S}_{3}^{(1)}\right) =-2s\tilde{S}_{4}^{(1)},  \label{r1} \\
&&\left( \tilde{S}_{1}^{(1)},\tilde{S}_{3}^{(4)}\right) +\left( \tilde{S}%
_{1}^{(4)},\tilde{S}_{3}^{(1)}\right) +\left( \tilde{S}_{2}^{(1)},\tilde{S}%
_{2}^{(4)}\right) =-s\tilde{S}_{4}^{(4)},  \label{r2} \\
&&2\left( \tilde{S}_{1}^{(1)},\tilde{S}_{3}^{(2)}\right) +2\left( \tilde{S}%
_{1}^{(2)},\tilde{S}_{3}^{(1)}\right) +2\left( \tilde{S}_{1}^{(2)},\tilde{S}%
_{3}^{(2)}\right)  \notag \\
&&+2\left( \tilde{S}_{2}^{(1)},\tilde{S}_{2}^{(2)}\right) +\left( \tilde{S}%
_{2}^{(2)},\tilde{S}_{2}^{(2)}\right) =-2s\tilde{S}_{4}^{(2)},  \label{r3} \\
&&\left( \tilde{S}_{1}^{(1)},\tilde{S}_{3}^{(3)}\right) +\left( \tilde{S}%
_{1}^{(3)},\tilde{S}_{3}^{(1)}\right) +\left( \tilde{S}_{1}^{(2)},\tilde{S}%
_{3}^{(3)}\right)  \notag \\
&&+\left( \tilde{S}_{1}^{(3)},\tilde{S}_{3}^{(2)}\right) +\left( \tilde{S}%
_{2}^{(1)},\tilde{S}_{2}^{(3)}\right) +\left( \tilde{S}_{2}^{(2)},\tilde{S}%
_{2}^{(3)}\right) =-s\tilde{S}_{4}^{(3)},  \label{r4'}
\end{eqnarray}
where $\left( \tilde{S}_{4}^{(m)}\right) _{m=\overline{1,4}}$ represent the
components of the solution to the master equation for the theory with the
standard Lagrangian (\ref{1}) of order four in the coupling constant
\begin{equation}
\tilde{S}_{4}=\tilde{S}_{4}^{(1)}+\tilde{S}_{4}^{(2)}+\tilde{S}_{4}^{(3)}+%
\tilde{S}_{4}^{(4)},  \label{dezv4}
\end{equation}
i.e.
\begin{equation}
2s\tilde{S}_{4}+2\left( \tilde{S}_{1},\tilde{S}_{3}\right) +\left( \tilde{S}%
_{2},\tilde{S}_{2}\right) =0.  \label{un4b}
\end{equation}
The fourth-order deformation of the solution of the master equation
associated with the free theory (\ref{fract}), $S_{4}$, is solution to the
equation
\begin{equation}
2sS_{4}+2\left( S_{1},S_{3}\right) +\left( S_{2},S_{2}\right) =0,
\label{un4a}
\end{equation}
which results from (\ref{masts}) (with $S$ developed as in (\ref{our}))
projected on $\lambda ^{4}$. Subtracting (\ref{un4b}) from (\ref{un4a}) and
employing (\ref{un0}) and (\ref{un3}) we obtain
\begin{equation}
s\left( S_{4}-\tilde{S}_{4}\right) =-\left( \tilde{S}_{1},S_{3}-\tilde{S}%
_{3}\right) =-\left( \tilde{S}_{1},\sum\limits_{m=1}^{4}k_{3}^{(m)}\tilde{S}%
_{1}^{(m)}\right) .  \label{un5}
\end{equation}
Inserting (\ref{p1})--(\ref{p3}) in (\ref{un5}), we further deduce
\begin{eqnarray}
s\left( S_{4}-\tilde{S}_{4}\right) &=&s\left[ 2k_{3}^{(1)}\tilde{S}%
_{2}^{(1)}+\left( k_{3}^{(4)}+k_{3}^{(1)}\right) \tilde{S}_{2}^{(4)}+\left(
k_{3}^{(2)}+k_{3}^{(3)}\right) \tilde{S}_{2}^{(3)}\right]  \notag \\
&&+\left( k_{3}^{(2)}+k_{3}^{(1)}\right) \left( \tilde{S}_{1}^{(1)},\tilde{S}%
_{1}^{(2)}\right) +k_{3}^{(2)}\left( \tilde{S}_{1}^{(2)},\tilde{S}%
_{1}^{(2)}\right) .  \label{def4}
\end{eqnarray}
Taking into account the first relation from (\ref{p2}), it follows that the
right-hand side of (\ref{def4}) is $s$-exact if and only if the constants $%
k_{3}^{(2)}$ and $k_{3}^{(1)}$ from (\ref{un3}) are equal
\begin{equation}
k_{3}^{(2)}=k_{3}^{(1)}.  \label{eccons2}
\end{equation}
Substituting (\ref{eccons2}) in (\ref{un3}) we determine the general
expression of the third-order deformation of the fully deformed solution (%
\ref{our}) of the master equation associated with the free theory (\ref%
{fract}), $S_{3}$, in terms of some of the components of the solution to the
master equation for the theory with the standard Lagrangian (\ref{1}) under
the form
\begin{equation}
S_{3}=\tilde{S}_{3}+k_{3}^{(1)}\left( \tilde{S}_{1}^{(1)}+\tilde{S}%
_{1}^{(2)}\right) +k_{3}^{(3)}\tilde{S}_{1}^{(3)}+k_{3}^{(4)}\tilde{S}%
_{1}^{(4)}.  \label{un3f}
\end{equation}
Based on the same result, namely (\ref{eccons2}), from (\ref{def4}) we infer
the equation satisfied by the fourth-order deformation $S_{4}$
\begin{eqnarray}
&&s\left[ S_{4}-\tilde{S}_{4}-2k_{3}^{(1)}\tilde{S}_{2}^{(1)}-2k_{3}^{(1)}%
\tilde{S}_{2}^{(2)}-\left( k_{3}^{(1)}+k_{3}^{(3)}\right) \tilde{S}%
_{2}^{(3)}\right.  \notag \\
&&\quad \left. -\left( k_{3}^{(4)}+k_{3}^{(1)}\right) \tilde{S}_{2}^{(4)}%
\right] =0,  \label{eqs4}
\end{eqnarray}
which, according to the general result from subsection \ref{firstord} (see
also the argument leading to (\ref{un3})), possesses the solution
\begin{eqnarray}
S_{4}&=&\tilde{S}_{4}+2k_{3}^{(1)}\left( \tilde{S}_{2}^{(1)}+\tilde{S}%
_{2}^{(2)}\right) +\left( k_{3}^{(1)}+k_{3}^{(3)}\right) \tilde{S}_{2}^{(3)}
\notag \\
&&+\left( k_{3}^{(4)}+k_{3}^{(1)}\right) \tilde{S}_{2}^{(4)}+\sum%
\limits_{m=1}^{4}k_{4}^{(m)}\tilde{S}_{1}^{(m)},  \label{un6}
\end{eqnarray}
where $\left( k_{4}^{(m)}\right) _{m=\overline{1,4}}$ are some arbitrary,
real constants. This ends the first step of the uniqueness procedure.

Next, we proceed like we did in the above for $S_{3}$ and $S_{4}$, but in
relation with $S_{4}$ and $S_{5}$. Inserting expansions (\ref{ein}) and (\ref%
{our}) respectively into equations (\ref{maststilde}) and (\ref{masts})
projected on $\lambda ^{5}$, we find the equations satisfied by $S_{5}$ and $%
\tilde{S}_{5}$ respectively under the form
\begin{eqnarray}
sS_{5}+\left( S_{1},S_{4}\right) +\left( S_{2},S_{3}\right) &=&0,
\label{un7} \\
s\tilde{S}_{5}+\left( \tilde{S}_{1},\tilde{S}_{4}\right) +\left( \tilde{S}%
_{2},\tilde{S}_{3}\right) &=&0.  \label{un8}
\end{eqnarray}
If we subtract (\ref{un8}) from (\ref{un7}) and recall (\ref{un0}), then we
infer the equation
\begin{equation}
s\left( S_{5}-\tilde{S}_{5}\right) =-\left( \tilde{S}_{1},S_{4}-\tilde{S}%
_{4}\right) -\left( \tilde{S}_{2},S_{3}-\tilde{S}_{3}\right) .  \label{un9}
\end{equation}
By replacing (\ref{un3f}) and (\ref{un6}) into the right-hand side of (\ref%
{un9}) and by further calculating the resulting expression with the help of
relations (\ref{p1})--(\ref{p3}) and (\ref{q1})--(\ref{q4}), we arrive at
\begin{eqnarray}
s\left( S_{5}-\tilde{S}_{5}\right) &=&s\left[ 2k_{4}^{(1)}\tilde{S}%
_{2}^{(1)}+\left( k_{4}^{(4)}+k_{4}^{(1)}\right) \tilde{S}_{2}^{(4)}+\left(
k_{4}^{(2)}+k_{4}^{(3)}\right) \tilde{S}_{2}^{(3)}\right.  \notag \\
&&+3k_{3}^{(1)}\left( \tilde{S}_{3}^{(1)}+\tilde{S}_{3}^{(2)}\right) +\left(
2k_{3}^{(1)}+k_{3}^{(3)}\right) \tilde{S}_{3}^{(3)}  \notag \\
&&\left. +\left( 2k_{3}^{(1)}+k_{3}^{(4)}\right) \tilde{S}_{3}^{(4)}\right]
+\left( k_{4}^{(2)}+k_{4}^{(1)}\right) \left( \tilde{S}_{1}^{(1)},\tilde{S}%
_{1}^{(2)}\right)  \notag \\
&&+k_{4}^{(2)}\left( \tilde{S}_{1}^{(2)},\tilde{S}_{1}^{(2)}\right) .
\label{def5}
\end{eqnarray}
The last equation demands that the right-hand side of (\ref{def5}) is $s$%
-exact. Due to the first equation from (\ref{p2}), this is attained if and
only if the constants $k_{4}^{(2)}$ and $k_{4}^{(1)}$ are equal
\begin{equation}
k_{4}^{(2)}=k_{4}^{(1)}.  \label{eccons3}
\end{equation}
Substituting (\ref{eccons3}) back in (\ref{un6}) and (\ref{def5})
respectively, on the one hand we deduce the general form of the fourth-order
deformation of the fully deformed solution (\ref{our}) of the master
equation associated with the free theory (\ref{fract}), $S_{4}$, in terms of
some of the components of the solution to the master equation for the theory
with the standard Lagrangian (\ref{1})
\begin{eqnarray}
S_{4} &=&\tilde{S}_{4}+2k_{3}^{(1)}\left( \tilde{S}_{2}^{(1)}+\tilde{S}%
_{2}^{(2)}\right) +\left( k_{3}^{(1)}+k_{3}^{(3)}\right) \tilde{S}%
_{2}^{(3)}+\left( k_{3}^{(4)}+k_{3}^{(1)}\right) \tilde{S}_{2}^{(4)}  \notag
\\
&&+k_{4}^{(1)}\left( \tilde{S}_{1}^{(1)}+\tilde{S}_{1}^{(2)}\right)
+k_{4}^{(3)}\tilde{S}_{1}^{(3)}+k_{4}^{(4)}\tilde{S}_{1}^{(4)}  \label{un10}
\end{eqnarray}
and on the other hand we output the equation that must be fulfilled by the
fifth-order deformation $S_{5}$
\begin{eqnarray}
&&s\left[ S_{5}-\tilde{S}_{5}-3k_{3}^{(1)}\left( \tilde{S}_{3}^{(1)}+\tilde{S%
}_{3}^{(2)}\right) -\left( 2k_{3}^{(1)}+k_{3}^{(3)}\right) \tilde{S}%
_{3}^{(3)}\right.  \notag \\
&& -\left( 2k_{3}^{(1)}+k_{3}^{(4)}\right) \tilde{S}_{3}^{(4)}
-2k_{4}^{(1)}\left( \tilde{S}_{2}^{(1)}+\tilde{S}_{2}^{(2)}\right)  \notag \\
&&\left. -\left( k_{4}^{(4)}+k_{4}^{(1)}\right) \tilde{S}_{2}^{(4)}-\left(
k_{4}^{(1)}+k_{4}^{(3)}\right) \tilde{S}_{2}^{(3)}\right] =0.  \label{eqs5}
\end{eqnarray}
Using the same arguments like before it results that the general solution to
the last equation reads as
\begin{eqnarray}
S_{5}&=&\tilde{S}_{5}+3k_{3}^{(1)}\left( \tilde{S}_{3}^{(1)}+\tilde{S}%
_{3}^{(2)}\right) +\left( 2k_{3}^{(1)}+k_{3}^{(3)}\right) \tilde{S}_{3}^{(3)}
\notag \\
&& +\left( 2k_{3}^{(1)}+k_{3}^{(4)}\right) \tilde{S}_{3}^{(4)}
+2k_{4}^{(1)}\left( \tilde{S}_{2}^{(1)}+\tilde{S}_{2}^{(2)}\right)  \notag \\
&& +\left( k_{4}^{(4)}+k_{4}^{(1)}\right) \tilde{S}_{2}^{(4)}+\left(
k_{4}^{(1)}+k_{4}^{(3)}\right) \tilde{S}_{2}^{(3)}+\sum%
\limits_{m=1}^{4}k_{5}^{(m)}\tilde{S}_{1}^{(m)},  \label{un11}
\end{eqnarray}
with $\left( k_{5}^{(m)}\right) _{m=\overline{1,4}}$ some arbitrary, real
constants. This completes the second step of the uniqueness procedure.

We reprise the procedure used previously for $S_{4}$ and $S_{5}$, but in
connection with $S_{5}$ and $S_{6}$. In view of this, we project (\ref%
{maststilde}) and (\ref{masts}) on $\lambda ^{6}$, respectively, which
provides the equations
\begin{eqnarray}
2sS_{6}+2\left( S_{1},S_{5}\right) +2\left( S_{2},S_{4}\right) +\left(
S_{3},S_{3}\right) &=&0,  \label{un12} \\
2s\tilde{S}_{6}+2\left( \tilde{S}_{1},\tilde{S}_{5}\right) +2\left( \tilde{S}%
_{2},\tilde{S}_{4}\right) +\left( \tilde{S}_{3},\tilde{S}_{3}\right) &=&0,
\label{un13}
\end{eqnarray}
and then subtract the above relations one from the other and employ (\ref%
{un0}), obtaining
\begin{eqnarray}
2s\left( S_{6}-\tilde{S}_{6}\right) &=&-2\left( \tilde{S}_{1},S_{5}-\tilde{S}%
_{5}\right) -2\left( \tilde{S}_{2},S_{4}-\tilde{S}_{4}\right)  \notag \\
&&+\left( \tilde{S}_{3},\tilde{S}_{3}\right) -\left( S_{3},S_{3}\right) .
\label{un14}
\end{eqnarray}
Replacing (\ref{un3f}), (\ref{un10}), and (\ref{un11}) in the right-hand
side of (\ref{un14}) and further computing its expression by means of
relations (\ref{p1})--(\ref{p3}), (\ref{q1})--(\ref{q4}), and (\ref{r1})--(%
\ref{r4'}), we reach the equation
\begin{eqnarray}
2s\left( S_{6}-\tilde{S}_{6}\right) &=&s\left[ 8k_{3}^{(1)}\left( \tilde{S}%
_{4}^{(1)}+\tilde{S}_{4}^{(2)}\right) +2\left(
3k_{3}^{(1)}+k_{3}^{(3)}\right) \tilde{S}_{4}^{(3)}\right.  \notag \\
&&+2\left( 3k_{3}^{(1)}+k_{3}^{(4)}\right) \tilde{S}_{4}^{(4)}+6k_{4}^{(1)}%
\left( \tilde{S}_{3}^{(1)}+\tilde{S}_{3}^{(2)}\right)  \notag \\
&& +2\left( 2k_{4}^{(1)}+k_{4}^{(3)}\right) \tilde{S}_{3}^{(3)} +2\left(
2k_{4}^{(1)}+k_{4}^{(4)}\right) \tilde{S}_{3}^{(4)}  \notag \\
&&+2\left( 2k_{5}^{(1)}+\left( k_{3}^{(1)}\right) ^{2}\right) \tilde{S}%
_{2}^{(1)}+2\left( k_{3}^{(1)}\right) ^{2}\tilde{S}_{2}^{(2)}  \notag \\
&& +2\left( k_{5}^{(1)}+k_{5}^{(3)}+k_{3}^{(1)}k_{3}^{(3)}\right) \tilde{S}%
_{2}^{(3)}  \notag \\
&&\left. +2\left( k_{5}^{(1)}+k_{5}^{(4)}+k_{3}^{(1)}k_{3}^{(4)}\right)
\tilde{S}_{2}^{(4)}\right]  \notag \\
&&+2\left( k_{5}^{(2)}+k_{5}^{(1)}\right) \left( \tilde{S}_{1}^{(1)},\tilde{S%
}_{1}^{(2)}\right) +2k_{5}^{(2)}\left( \tilde{S}_{1}^{(2)},\tilde{S}%
_{1}^{(2)}\right) .  \label{def6}
\end{eqnarray}
On account of the former relation in (\ref{p2}), we conclude that (\ref{def6}%
) holds (i.e. its right-hand side is $s$-exact) if and only if the constants
$k_{4}^{(2)}$ and $k_{4}^{(1)}$ are equal
\begin{equation}
k_{5}^{(2)}=k_{5}^{(1)}.  \label{eccons4}
\end{equation}
Based on the last result inserted in (\ref{un11}) and (\ref{def6}), we
complete the third step of our procedure for constructing $S$ and in fact
proving the uniqueness of $\tilde{S}$: we output the general form of the
fifth-order deformation of the fully deformed solution (\ref{our}) of the
master equation associated with the free theory (\ref{fract})
\begin{eqnarray}
S_{5}&=&\tilde{S}_{5}+3k_{3}^{(1)}\left( \tilde{S}_{3}^{(1)}+\tilde{S}%
_{3}^{(2)}\right) +\left( 2k_{3}^{(1)}+k_{3}^{(3)}\right) \tilde{S}%
_{3}^{(3)}+\left( 2k_{3}^{(1)}+k_{3}^{(4)}\right) \tilde{S}_{3}^{(4)}  \notag
\\
&&+2k_{4}^{(1)}\left( \tilde{S}_{2}^{(1)}+\tilde{S}_{2}^{(2)}\right) +\left(
k_{4}^{(4)}+k_{4}^{(1)}\right) \tilde{S}_{2}^{(4)}+\left(
k_{4}^{(1)}+k_{4}^{(3)}\right) \tilde{S}_{2}^{(3)}  \notag \\
&&+k_{5}^{(1)}\left( \tilde{S}_{1}^{(1)}+\tilde{S}_{1}^{(2)}\right)
+k_{5}^{(3)}\tilde{S}_{1}^{(3)}+k_{5}^{(4)}\tilde{S}_{1}^{(4)}  \label{un15}
\end{eqnarray}
and meanwhile deduce the equation verified by the deformation of the next
order
\begin{eqnarray}
&&s\left[ S_{6}-\tilde{S}_{6}-4k_{3}^{(1)}\left( \tilde{S}_{4}^{(1)}+\tilde{S%
}_{4}^{(2)}\right) -\left( 3k_{3}^{(1)}+k_{3}^{(3)}\right) \tilde{S}%
_{4}^{(3)}\right.  \notag \\
&&-\left( 3k_{3}^{(1)}+k_{3}^{(4)}\right) \tilde{S}_{4}^{(4)}-3k_{4}^{(1)}%
\left( \tilde{S}_{3}^{(1)}+\tilde{S}_{3}^{(2)}\right) -\left(
2k_{4}^{(1)}+k_{4}^{(3)}\right) \tilde{S}_{3}^{(3)}  \notag \\
&&-\left( 2k_{4}^{(1)}+k_{4}^{(4)}\right) \tilde{S}_{3}^{(4)}-\left(
2k_{5}^{(1)}+\left( k_{3}^{(1)}\right) ^{2}\right) \left( \tilde{S}%
_{2}^{(1)}+\tilde{S}_{2}^{(2)}\right)  \notag \\
&&\left. -\left( k_{5}^{(1)}+k_{5}^{(3)}+k_{3}^{(1)}k_{3}^{(4)}\right)
\tilde{S}_{2}^{(3)}-\left(
k_{5}^{(1)}+k_{5}^{(4)}+k_{3}^{(1)}k_{3}^{(4)}\right) \tilde{S}_{2}^{(4)}%
\right] =0.  \label{eqs6}
\end{eqnarray}
The solution to this equation is written as
\begin{eqnarray}
S_{6}&=&\tilde{S}_{6}+4k_{3}^{(1)}\left( \tilde{S}_{4}^{(1)}+\tilde{S}%
_{4}^{(2)}\right) +\left( 3k_{3}^{(1)}+k_{3}^{(3)}\right) \tilde{S}_{4}^{(3)}
\notag \\
&&+\left( 3k_{3}^{(1)}+k_{3}^{(4)}\right) \tilde{S}_{4}^{(4)}+3k_{4}^{(1)}%
\left( \tilde{S}_{3}^{(1)}+\tilde{S}_{3}^{(2)}\right) +\left(
2k_{4}^{(1)}+k_{4}^{(3)}\right) \tilde{S}_{3}^{(3)}  \notag \\
&&+\left( 2k_{4}^{(1)}+k_{4}^{(4)}\right) \tilde{S}_{3}^{(4)}+\left(
2k_{5}^{(1)}+\left( k_{3}^{(1)}\right) ^{2}\right) \left( \tilde{S}%
_{2}^{(1)}+\tilde{S}_{2}^{(2)}\right)  \notag \\
&&+\left( k_{5}^{(1)}+k_{5}^{(3)}+k_{3}^{(1)}k_{3}^{(3)}\right) \tilde{S}%
_{2}^{(3)}+\left( k_{5}^{(1)}+k_{5}^{(4)}+k_{3}^{(1)}k_{3}^{(4)}\right)
\tilde{S}_{2}^{(4)}  \notag \\
&&+\sum\limits_{m=1}^{4}k_{6}^{(m)}\tilde{S}_{1}^{(m)},  \label{un16}
\end{eqnarray}
with $\left( k_{6}^{(m)}\right) _{m=\overline{1,4}}$ some arbitrary, real
constants, independent so far. Just like in the above it can be shown that
in fact $k_{6}^{(2)}=k_{6}^{(1)}$ (via establishing a relationship between $%
S_{7}$ and $\tilde{S}_{7}$), etc.

Replacing (\ref{un0}), (\ref{un3f}), (\ref{un10}), and (\ref{un15}) (for $%
k_{6}^{(2)}=k_{6}^{(1)}$) in (\ref{our}) and regrouping the various terms
according to the structure of decompositions (\ref{dezv1}), (\ref{dezv2}), (%
\ref{dezv3}), (\ref{dezv4}), we finally obtain
\begin{eqnarray*}
&&S=\tilde{S}_{0}+\lambda \left( 1+k_{3}^{(1)}\lambda
^{2}+k_{4}^{(1)}\lambda ^{3}+k_{5}^{(1)}\lambda ^{4}+k_{6}^{(1)}\lambda
^{5}+\cdots \right) \left( \tilde{S}_{1}^{(1)}+\tilde{S}_{1}^{(2)}\right) \\
&&+\lambda \left( 1+k_{3}^{(3)}\lambda ^{2}+k_{4}^{(3)}\lambda
^{3}+k_{5}^{(3)}\lambda ^{4}+k_{6}^{(3)}\lambda ^{5}+\cdots \right) \tilde{S}%
_{1}^{(3)} \\
&&+\lambda \left( 1+k_{3}^{(4)}\lambda ^{2}+k_{4}^{(4)}\lambda
^{3}+k_{5}^{(4)}\lambda ^{4}+k_{6}^{(4)}\lambda ^{5}+\cdots \right) \tilde{S}%
_{1}^{(4)} \\
&&+\lambda ^{2}\left[ 1+2k_{3}^{(1)}\lambda ^{2}+2k_{4}^{(1)}\lambda
^{3}+\left( 2k_{5}^{(1)}+\left( k_{3}^{(1)}\right) ^{2}\right) \lambda
^{4}+\cdots \right] \left( \tilde{S}_{2}^{(1)}+\tilde{S}_{2}^{(2)}\right) \\
&&+\lambda ^{2}\left[ 1+\left( k_{3}^{(1)}+k_{3}^{(3)}\right) \lambda
^{2}+\left( k_{4}^{(1)}+k_{4}^{(3)}\right) \lambda ^{3}\right. \\
&&\left. +\left( k_{5}^{(1)}+k_{5}^{(3)}+k_{3}^{(1)}k_{3}^{(3)}\right)
\lambda ^{4}+\cdots \right] \tilde{S}_{2}^{(3)} \\
&&+\lambda ^{2}\left[ 1+\left( k_{3}^{(1)}+k_{3}^{(4)}\right) \lambda
^{2}+\left( k_{4}^{(1)}+k_{4}^{(4)}\right) \lambda ^{3}\right. \\
&&\left. +\left( k_{5}^{(1)}+k_{5}^{(4)}+k_{3}^{(1)}k_{3}^{(4)}\right)
\lambda ^{4}+\cdots \right] \tilde{S}_{2}^{(4)} \\
&&+\lambda ^{3}\left( 1+3k_{3}^{(1)}\lambda ^{2}+3k_{4}^{(1)}\lambda
^{3}+\cdots \right) \left( \tilde{S}_{3}^{(1)}+\tilde{S}_{3}^{(2)}\right) \\
&&+\lambda ^{3}\left[ 1+\left( 2k_{3}^{(1)}+k_{3}^{(3)}\right) \lambda
^{2}+\left( 2k_{4}^{(1)}+k_{4}^{(3)}\right) \lambda ^{3}+\cdots \right]
\tilde{S}_{3}^{(3)} \\
&&+\lambda ^{3}\left[ 1+\left( 2k_{3}^{(1)}+k_{3}^{(4)}\right) \lambda
^{2}+\left( 2k_{4}^{(1)}+k_{4}^{(4)}\right) \lambda ^{3}+\cdots \right]
\tilde{S}_{3}^{(4)} \\
&&+\lambda ^{4}\left( 1+4k_{3}^{(1)}\lambda ^{2}+\cdots \right) \left(
\tilde{S}_{4}^{(1)}+\tilde{S}_{4}^{(2)}\right) \\
&&+\lambda ^{4}\left[ 1+\left( 3k_{3}^{(1)}+k_{3}^{(3)}\right) \lambda
^{2}+\cdots \right] \tilde{S}_{4}^{(3)} \\
&&+\lambda ^{4}\left[ 1+\left( 3k_{3}^{(1)}+k_{3}^{(4)}\right) \lambda
^{2}+\cdots \right] \tilde{S}_{4}^{(4)}+\cdots .
\end{eqnarray*}%
It is now clear that the last expression can be written as in (\ref{ein})
(at least in the first orders in $\lambda $) modulo the transformations (\ref%
{deflambda})--(\ref{defq}). The conclusion of this section is that the
deformation procedure for action (\ref{1}) can be used at proving in an
elegant manner the uniqueness of eleven-dimensional interactions between a
graviton and a three-form gauge field prescribed by General Relativity.

\section{Conclusion\label{conc}}

To conclude with, in this paper we have generated the consistent
interactions in eleven spacetime dimensions that can be added to a free
theory describing a massless spin-two field and an Abelian three-form gauge
field. Our treatment is based on the Lagrangian BRST deformation procedure,
which relies on the construction of consistent deformations of the solution
to the master equation with the help of standard cohomological techniques.
The couplings are obtained under the hypotheses of smoothness in the
coupling constant, locality, Lorentz covariance, Poincar\'{e} invariance,
and the presence of at most two derivatives in the interacting Lagrangian.
Our main result is that if we decompose the metric like $g_{\mu \nu }=\sigma
_{\mu \nu }+\lambda h_{\mu \nu }$, then we can couple the Abelian three-form
gauge field to $h_{\mu \nu }$ in the space of formal series with the maximum
derivative order equal to two in $h_{\mu \nu }$ such that the resulting
interactions agree with the usual couplings between the three-form and the
massless spin-two field in vielbein formulation. Thus, we emphasize the
uniqueness of eleven-dimensional interactions between a graviton and a
three-form gauge field prescribed by General Relativity. We cannot stress
enough that the cosmological term is not restricted in this context. Its
presence is forbidden only if we add to the present field content other
particles, such as massless gravitini.

\section*{Acknowledgments}

The authors wish to thank Constantin Bizdadea and Odile Saliu for useful
discussions and comments. This work is partially supported by the European
Commission FP6 program MRTN-CT-2004-005104 and by the grant AT24/2005 with
the Romanian National Council for Academic Scientific Research
(C.N.C.S.I.S.) and the Romanian Ministry of Education and Research (M.E.C.).

\end{document}